\documentclass[12pt,twoside]{article}
\usepackage[mode=buildnew]{standalone}
\usepackage{graphicx}
\usepackage[natbib=true,
backend=biber,
maxnames=6,
minnames=3,
maxcitenames=2,
mincitenames=1,
style=ama,
autocite=superscript,
giveninits=true,
uniquename=false,
uniquelist=false,
bibencoding=utf8]{biblatex}
\usepackage{booktabs}
\usepackage{array}
\usepackage[font=footnotesize,labelfont=bf]{caption}
\usepackage{subcaption}
\usepackage{hyperref}
\hypersetup{
    colorlinks=true,
    citecolor=egyptianblue,
    linkcolor=darkcoral,   
    filecolor=black,
    urlcolor=darkolivegreen,
    bookmarksopen=true,
    bookmarksopenlevel=3,
    plainpages=false,
    pdftitle={Direct optimization of the probability of lesion origin in proton treatment planning for low-grade glioma patients},
    pdfpagemode=FullScreen,
    }
\usepackage[all]{hypcap}
\usepackage{siunitx}
\DeclareSIUnit{\pp}{\textup{p.p.}}
\usepackage{amsmath, amsxtra, amstext, amssymb, amsthm, amsfonts}
\usepackage{dsfont}
\usepackage{mathtools}
\usepackage{tikz}
\usepackage{multirow}
\usepackage{bigstrut}
\usepackage{authblk}
\usepackage{abstract}
\usepackage[mathlines]{lineno}
\usepackage{float}
\usepackage{fancyhdr}
\usepackage{color}
\usepackage{csquotes}
\usepackage[capitalise]{cleveref}
\usepackage{orcidlink}

\newcolumntype{C}[1]{>{\centering\let\newline\\\arraybackslash\hspace{0pt}}m{#1}}
\newcolumntype{M}[1]{>{\centering\arraybackslash}m{#1}}

\definecolor{egyptianblue}{rgb}{0.06, 0.2, 0.65}
\definecolor{darkolivegreen}{rgb}{0.33, 0.42, 0.18}
\definecolor{darkcoral}{rgb}{0.8, 0.36, 0.27}
\definecolor{myteal}{rgb}{0.0196, 0.60, 0.549}
\definecolor{denim}{rgb}{0.08, 0.38, 0.74}

\makeatletter \renewcommand\@biblabel[1]{$^{#1}$} \makeatother
\begin{document}

\newcommand{\solid}{\raisebox{2pt}{\tikz{\draw[black,solid,line width = 0.9pt](0,0) -- (4mm,0);}}}
\newcommand{\dashdot}{\raisebox{2pt}{\tikz{\draw[black,dash dot,line width = 0.9pt](0,0) -- (4mm,0);}}}
\newcommand{\dashed}{\raisebox{2pt}{\tikz{\draw[black,dashed,line width = 0.9pt](0,0) -- (4mm,0);}}}
\newcommand{\dotted}{\raisebox{2pt}{\tikz{\draw[black,dotted,line width = 0.9pt](0,0) -- (4mm,0);}}}

\pagestyle{fancy}
\renewcommand{\sectionmark}[1]{\markboth{#1}{}}
\fancyhead{}
\renewcommand{\headrulewidth}{0.5pt}%
\renewcommand{\headrule}{\hbox to\headwidth{%
  \color{darkgray}\leaders\hrule height \headrulewidth\hfill}}
\fancyhead[RE]{\small\color{darkgray}\thepage \vspace{-10pt}}
\fancyhead[LE]{\small\color{darkgray} Ortkamp et al. (2026): Direct optimization of the probability of lesion origin \vspace{-10pt}}
\fancyhead[RO]{\small\color{darkgray}\nouppercase{\leftmark} \vspace{-10pt}}
\fancyhead[LO]{\small\color{darkgray}\thepage \vspace{-10pt}}

\fancyfoot{}
\renewcommand{\footrulewidth}{0.5pt}%
\renewcommand{\footrule}{\hbox to\headwidth{%
  \color{darkgray}\leaders\hrule height \footrulewidth\hfill}}
\fancyfoot[RO, LE]{\small\color{darkgray} Last edited \date{\today}}

\setlength{\headheight}{19.69997pt}

\title{Direct optimization of the probability of lesion origin in proton treatment planning for low-grade glioma patients}


\renewcommand\Authfont{\normalsize}
\renewcommand\Affilfont{\itshape\footnotesize}

\author[1,2,3]{Tim Ortkamp\thanks{Corresponding author: Tim Ortkamp, Scientific Computing Center, Karlsruhe Institute of Technology (KIT), Hermann-von-Helmholtz-Platz 1, 76344 Eggenstein-Leopoldshafen, Email: \href{tim.ortkamp@kit.edu}{tim.ortkamp@kit.edu}}}
\author[4,5]{Habiba Sallem}
\author[4,5,6]{Semi Harrabi}
\author[1,3]{Martin Frank}
\author[2,3,5,6]{Oliver Jäkel}
\author[4,5,7]{Julia Bauer}
\author[2,5]{Niklas Wahl}

\affil[1]{Karlsruhe Institute of Technology (KIT), Scientific Computing Center, Hermann-von-Helmholtz-Platz 1, 76344 Eggenstein-Leopoldshafen, Germany}
\affil[2]{Department of Medical Physics in Radiation Oncology, German Cancer Research Center (DKFZ), 69120 Heidelberg, Germany}
\affil[3]{Helmholtz Information and Data Science School for Health (HIDSS4Health), Karlsruhe/Heidelberg, Germany}
\affil[4]{Department of Radiation Oncology, University Hospital Heidelberg, 69120 Heidelberg, Germany}
\affil[5]{National Center for Radiation Research in Oncology (NCRO), Heidelberg Institute for Radiation Oncology (HIRO), 69120 Heidelberg, Germany}
\affil[6]{Heidelberg Ion Beam Therapy Center (HIT), 69120 Heidelberg, Germany}
\affil[7]{Department of Radiation Oncology, Charité - Universitätsmedizin Berlin, Corporate Member of Freie Universität Berlin and Humboldt-Universität zu Berlin, 13353 Berlin, Germany}

\date{\normalsize Version typeset \today \\[1em] \begin{minipage}{\textwidth}\footnotesize{This is the peer reviewed version of the following article: \textit{\enquote{Ortkamp T, Sallem H, Harrabi S, et al. Direct optimization of the probability of lesion origin in proton treatment planning for low-grade glioma patients. Med Phys. 2026;53:e70395.}}, which has been published in final form at \href{https://doi.org/10.1002/mp.70395}{DOI: 10.1002/mp.70395}.}\end{minipage}}

\maketitle
\thispagestyle{empty}
\renewcommand{\thefootnote}{\arabic{footnote}}


\begin{abstract}
\noindent\textbf{Background:} In proton therapy of low-grade glioma (LGG) patients, contrast-enhancing brain lesions (CEBLs) on magnetic resonance imaging are considered predictive of late radiation-induced lesions. From the observation that CEBLs tend to concentrate in regions of increased dose-averaged linear energy transfer (LET\textsubscript{d}) and proximal to the ventricular system, the probability of lesion origin (POLO) model has been established as a multivariate logistic regression model for the voxel-wise probability prediction of the CEBL origin.

\noindent\textbf{Purpose:} To date, leveraging the predictive power of the POLO model for treatment planning relies on hand tuning the dose and LET\textsubscript{d} distribution to minimize the resulting probability predictions. In this paper, we therefore propose automated POLO model-based treatment planning by directly integrating POLO calculation and optimization into plan optimization for LGG patients.

\noindent\textbf{Approach:} We introduce an extension of the original POLO model including a volumetric correction factor, and a model-based optimization scheme featuring a linear reformulation of the model together with feasible optimization functions based on the predicted POLO values. The developed framework is implemented in the open-source treatment planning toolkit matRad.

\noindent\textbf{Results:} Our framework can generate clinically acceptable treatment plans while automatically taking into account outcome predictions from the POLO model. It also supports the definition of customized POLO model-based objective and constraint functions. Optimization results from a sample LGG patient show that the POLO model-based outcome predictions can be minimized under expectable shifts in dose, LET\textsubscript{d}, and POLO distributions, while sustaining target coverage ($\Delta_{\text{PTV}} \text{D95}_{RBE,fx}\approx{0.00}$, $\Delta_{\text{GTV}} \text{D95}_{RBE,fx}\approx{0.03}$), even when NTCP is strongly down-regulated.

\noindent\textbf{Conclusion:} POLO model-based treatment plan optimization for LGG patients can be implemented in a technically feasible way, alleviating the need to hand tune the dose and LET\textsubscript{d} distribution. Future work should address multi-patient follow-up studies.
\vspace{0pt}
\end{abstract}


\section{Introduction} \label{sec:introduction}
Late radiation-induced contrast-enhancing brain lesions (CEBLs) pose an adverse event after proton therapy of brain tumors and, by tendency, cluster in regions of increased dose-averaged linear energy transfer and near the ventricular system (VS) \cite{Harrabi2022}. They are considered predictive of late radiation-induced lesions (LRLs), i.e., brain tissue that shows damage from radiation exposure with observable symptoms delayed for months to years after treatment. For the patients, LRLs can -- in rare cases -- severely impact quality of life through anatomical and functional impairments, such as vascular abnormalities, demyelination, gliosis, white matter necrosis, and cognitive impairment \cite{Greene2012}. Therefore, attention is paid to the prevention of the occurrence of post-treatment LRLs in patients undergoing proton therapy.

Backed by clinical evidence on CEBLs, the probability of lesion origin (POLO) model has been developed \cite{Bahn2020}, a voxel-level model that accurately predicts the localization of CEBLs in low-grade glioma (LGG) patients and allows mapping to a patient-level normal tissue complication probability (NTCP). The POLO model constitutes a multivariate
logistic regression function fitted on a cohort of 110 patients with LGG treated with proton therapy at the Heidelberg Ion-Beam Therapy Center (HIT). It incorporates three key risk factors: the voxel-wise total dose $d$, the dose-averaged linear energy transfer (LET\textsubscript{d}), denoted by $l_{d}$, and a binary risk factor $b$ measuring the proximity to the cerebral ventricles. Although the POLO model shows reasonably high predictive performance (AUC $\approx$ 0.94, Brier score $\approx$ 2.61 $\times$ 10$^{-5}$), there is only a limited amount of preceeding and follow-up research to date promoting its clinical implementation\cite{Eulitz2019,Bahn2020,Bauer2021,Harrabi2022,Eulitz2023,Sallem2024,Palkowitsch2025}.

Related work includes a series of publications on \emph{voxel-based outcome analysis} (VBA), which takes the full spatial dose distribution as input to detect patterns between local dose and treatment outcome \cite{Palma2020,Ebert2021,McWilliam2023} and has already been applied to different sites \cite{Acosta2013,Palma2016,Monti2017}. VBA establishes as a retrospective tool to perform statistical inference on the dose-response relationship, i.e., to generate \enquote{significance maps} from voxel-wise hypothesis testing or generalized linear models. However, McWilliam et al. \cite{McWilliam2023} even states that \enquote{the future direction of VBA is towards optimizing plans directly on the dose response maps}. This is accompanied by developments in the field of \emph{outcome-based treatment planning} \cite{Brahme1999,Semenenko2008,Kierkels2014,Ajdari2022,Maragno2024}, leveraging predictive models from radiobiology and machine learning to directly optimize treatment plans. These models are largely based on extracted dose features rather than the full spatial distribution. Hence, by integrating the POLO model into treatment plan optimization, we position ourselves in the current gap between these two research fields.

Regarding the POLO model, a previous study by Bauer et al. \cite{Bauer2021} has already explored the impact of various risk factors, patient characteristics, and treatment plan parameters on the POLO model-based risk prediction, acknowledging the need for \enquote{novel biologically guided plan optimization approaches to achieve effective risk minimization in proton therapy for LGG}. As a response, Sallem et al. \cite{Sallem2024} developed a risk-minimizing treatment planning optimization strategy, which incorporates a variable LET\textsubscript{d}-dependent proton RBE and actively spares the peri-ventricular region. This strategy is currently being applied in the world’s first prospective, randomized phase II trial (\href{https://clinicaltrials.gov/}{ClinicalTrials.gov} Identifier: NCT05964569) to evaluate the feasibility of POLO model-based risk minimization in clinical practice. However, this does not yet feature an automated optimization pipeline to directly integrate the POLO model into treatment planning.

In this paper, we leverage the clinical insights gained from Sallem et al. \cite{Sallem2024} to address the challenges arising from outcome model integration by introducing a framework for direct optimization of the POLO in intensity-modulated proton therapy planning for LGG patients. To this end, we implemented a forward calculation step to determine the spatial POLO distribution (POLO \enquote{map}) together with feasible POLO model-based optimization functions -- either based on the original POLO model or on a linear reformulation -- and derived the corresponding backward calculation step to yield the fluence gradient. The optimization functions then serve as independent quantities for treatment plan optimization.

We implemented our approach within the open-source treatment planning toolkit matRad \cite{Wieser2017}, using the nonlinear quasi-Newton interior-point methods from IPOPT \cite{Waechter2006} to minimize a weighted sum of dose-volume and POLO model-based objectives. While we aim to approach the quality of clinical reference plans, our focus here is on showing the feasibility and overall efficacy of POLO model-based optimization.

The paper is structured as follows: in \cref{sec:methods} we introduce the theoretical and application-related concepts underlying the integration framework, including a definition of the POLO model, the derived model-based optimization functions, their implementation and finally, some details on the design choices made for treatment planning. Results on the dose/LET\textsubscript{d} distributions, POLO maps, as well as treatment outcome predictions are then covered in \cref{sec:results}. We round off the paper with a discussion and concluding remarks in \cref{sec:discussion} and \cref{sec:conclusion}. For interested readers, \hyperref[sec:matrad-functions]{\textbf{Appendix}} includes mathematical definitions of the dose-volume functions used for conventional plan optimization, the computational costs for all optimization cases, and an example of robust optimization.


\section{Materials and Methods} \label{sec:methods}

\subsection{Voxel-wise POLO model} \label{subsec:polo}
The POLO model assigns each voxel from the patient geometry a local risk estimate between 0 and 1, indicating the probability of a voxel being the origin of a CEBL. Its original definition proposed by Bahn et al. \cite{Bahn2020} considers the voxel-wise total dose $d$ and its elementary product with the dose-averaged LET $l_{d}$ as physical inputs. We further denote the number of voxels by $n$, i.e., $d\in{\mathbb{R}^{n}_{+}}$ and $l_{d}\in{\mathbb{R}^{n}_{+}}$. 

\subsubsection{Definition}\label{subsubsec:polo-definition}
Under the assumptions for a multivariate logistic regression, the \emph{probability function} with the linear predictor coefficients taken from Bahn et al. \cite{Bahn2020} reads
\begin{align}\label{eq:orig-polo}
    p(\eta) = \sigma(\eta) = \frac{1}{1+\exp{(-\eta)}} \in (0,1)^{n}, \quad \eta=-26.3 + \beta_{1}\cdot{d} + \beta_{2}\cdot{(d\circ{l_{d}})} + 1.19\cdot{b}\,,
\end{align}
where $\beta_{1}=\SI{0.19}{\gray^{-1}}$, $\beta_{2}=\SI{0.018}{\gray^{-1}\keV\per\mu\meter^{-1}}$, and the binary risk factor
\begin{align}\label{eq:brf}
    b = \begin{pmatrix} \mathds{1}\{\delta_{1}\leq{4}\} & \cdots & \mathds{1}\{\delta_{n}\leq{4}\} \end{pmatrix}^{T}\in{\{0,1\}^{n}} \quad |\, \delta_{i}: \text{distance [\si{\milli\meter}] of $i$-th voxel to VS}
\end{align}
captures the proximity effect of the local risk in a radius around the VS.

\subsubsection{Extension with volume dependence}\label{subsubsec:polo-extension}
One effect that is not taken into account in the above definition is the volume dependence of the local risk estimates.
This may be considered in practice because the model parameters from \Cref{eq:orig-polo} are fitted on a planning CT with a spatial resolution of $0.6\times{0.6}\times{3.0}$ \si{\milli\meter\cubed}, while treatment planning usually takes place at a lower resolution. We argue that $p(\eta)$ could be adjusted based on the critical element model\cite{Niemierko1991} by expanding \Cref{eq:orig-polo} to
\begin{align}\label{eq:ext-polo}
    p(\eta, k) = 1-\left[1-p(\eta)\right]^{k} = 1-\frac{1}{\left[1+\exp{(\eta)}\right]^{k}} \in (0,1)^{n}\,,
\end{align}
including the volumetric correction factor
\begin{align}
    k = v_{\text{new}}/v_{\text{old}}\in{\mathbb{R}_{+}} \quad |\, v: \text{volume [\si{\milli\meter\cubed}] of a voxel}
\end{align}
which accounts for the rescaling of the local risk in the event of volumetric changes. In \Cref{eq:ext-polo}, we use the point symmetry of the sigmoid function that $1-p(\eta)=p(-\eta)$, and therefore we describe the volume-corrected probability of lesion origin $p(\eta, k)$ as the counter-probability of the exponentiated probability of \underline{no} lesion origin $p(-\eta)^{k}$. For example, if $p(\eta)_{i}=0.1$ for the $i$-th voxel, and we halve the spatial resolution, i.e., $k=2$, then the volume-corrected probability of lesion origin will be $p(\eta, k)_{i}=1-(1-0.1)^{2}=0.19$. This effectively reduces biases due to volumetric differences.

\subsubsection{Patient-level NTCP}\label{subsubsec:global-polo-ntcp}
Under the assumption of a serial tissue response, Bahn et al. \cite{Bahn2020} also derived a patient-level NTCP prediction function, which maps the voxel-wise POLO values $p(\eta, k)_{i}$ to a risk estimate of developing at least one CEBL:
\begin{align}\label{eq:ntcp}
   NTCP = 1 - \prod_{i=1}^{n}{(1-p(\eta, k)_{i})} \in(0,1)\,.
\end{align}

\subsection{POLO model-based optimization} \label{subsec:functions}
Our approach strives to directly optimize intensity-modulated proton treatment plans for LGG patients such that the voxel- and patient-level risk estimates returned by the POLO model are minimized while satisfying conventional dosimetric treatment goals. This necessitates the design of a scalar optimization function $f\colon{D}\rightarrow\mathbb{R}$, $D\subseteq{\mathbb{R}^{n}}$, and (ideally) the derivation of a corresponding vector-valued gradient function $\nabla_{x}f\colon{D}\rightarrow\mathbb{R}^{n}$, $x\in{D}$, to feed into a large-scale non-linear optimization algorithm, as well as the establishment of a gradient propagation rule from the model to the fluence (decision variable). The function $f$ serves as a generic operator that, when applied to the probability function $p$, takes the form $f_{p}\colon(0,1)^{n}\rightarrow\mathbb{R}$ with associated gradient $\nabla_{p}f_{p}\colon(0,1)^{n}\rightarrow\mathbb{R}^{n}$.

Let $\phi\in{\mathbb{R}_{+}^{m}}$ be the fluence vector whose dimensionality equals the number of proton beamlets $m$. Both $f_{p}$ and $\nabla_{\phi}f_{p}$ rely on the calculation of the product between $d$ and $l_{d}$. In accordance with Unkelbach et al.\cite{Unkelbach2016}, we define the $i$-th element of the quantity $(d\circ{l_{d}})$ as
\begin{align}\label{eq:dose-avg-let}
    d_{i}\cdot{l_{d,i}} = d_{i}\cdot\frac{\sum_{j}{\mathcal{L}_{ij}\mathcal{D}_{ij}\phi_{j}}}{\sum_{j}{\mathcal{D}_{ij}\phi_{j}}}=d_{i}\cdot\frac{1}{d_{i}}\sum_{j}{\underbrace{\mathcal{L}_{ij}\mathcal{D}_{ij}}_{\mathcal{M}_{ij}}\phi_{j}} = \sum_{j}{\mathcal{M}_{ij}\phi_{j}}\,,
\end{align}
where $i$ and $j$ index the voxels and proton spots. \Cref{eq:dose-avg-let} involves the fluence-independent LET- and dose-influence matrices $\mathcal{L}\in{\mathbb{R}}_{+}^{n\times{m}}$ and $\mathcal{D}\in{\mathbb{R}}_{+}^{n\times{m}}$, whose elementwise product $\mathcal{M}$ can be precomputed. While the forward calculation of $f_{p}$ is straightforward using \Cref{eq:dose-avg-let}, there are implications for the backward calculation of $\nabla_{\phi}f_{p}$. Before we introduce sample functions for $f_{p}$, we therefore like to state a general notation of the fluence gradient.

\subsubsection{Fluence gradient} 
Using basic calculus, the fluence gradient of a differentiable function $f_{p}$ reads
\begin{align}\label{eq:phi-grad}
    \nabla_{\phi}{f_{p}} &= \left[\nabla_{d}{\eta}\cdot\mathcal{J}_{d}{(\phi)}+\nabla_{(d\circ{l_{d}})}{\eta}\cdot\mathcal{J}_{(d\circ{l_{d}})}{(\phi)}\right]^{T}\left(\nabla_{\eta}{p}\circ\nabla_{p}{f_{p}}\right) = \mathcal{C}^{T}\left(\nabla_{\eta}{p}\circ\nabla_{p}{f_{p}}\right)\,,
\end{align}
under (i) $\nabla_{d}{\eta}=\beta_{1}$, $\nabla_{(d\circ{l_{d}})}{\eta}=\beta_{2}$, (ii) $\mathcal{J}_{d}{(\phi)}=\mathcal{D}$, (iii) as per \Cref{eq:dose-avg-let}, $\mathcal{J}_{(d\circ{l_{d}})}{(\phi)}=\mathcal{M}$, and (iv) $\beta_{1}\cdot\mathcal{D}+\beta_{2}\cdot\mathcal{M}\eqqcolon{\mathcal{C}}\in{\mathbb{R}_{+}^{n\times{m}}}$. We observe that the right-hand side of \Cref{eq:phi-grad} only depends on the analytical gradient $\nabla_{\eta}{p}$ and the specific structure of $\nabla_{p}{f_{p}}$.

\subsubsection{Linear reformulation}\label{subsubsec:convex-polo}
Regardless of the function $f_{p}$, integrating the probability function $p$ into proton treatment planning has two potential weak points: the (partial) non-convexity of the probability function $p$ from \Cref{eq:ext-polo} in the linear predictor $\eta$ and the flattening of $\nabla_{\eta}{p}$ towards the boundaries of $\eta$ and $k$. This implies that gradient-based optimizers may terminate at locally but not globally optimal points and/or be prevented from making sufficient steps towards smaller local risk estimates.

To overcome the disadvantageous properties of the function $p$, we propose a linear reformulation by exploiting its monotonicity. Since monotonic transformations preserve the order, we can simply invert both the sigmoid function $\sigma$ and the volumetric correction factor $k$ without changing the minimum point of $p$, yielding the linear \emph{decision function}
\begin{align}\label{eq:polo-surr}
    \tilde{p}(\eta) = \eta\in{[-26.3, \infty)}, \quad \nabla_{\eta}{\tilde{p}} = \mathbf{1}\,.
\end{align}
Accordingly, we define $f_{\tilde{p}}\colon\mathbb{R}^{n}\rightarrow\mathbb{R}$ and $\nabla_{\tilde{p}}f_{\tilde{p}}\colon\mathbb{R}^{n}\rightarrow\mathbb{R}^{n}$ on the linearized quantity $\tilde{p}$, and adapt the fluence gradient from \Cref{eq:phi-grad} to
\begin{align}\label{eq:phi-grad-surr}
    \nabla_{\phi}{f_{\tilde{p}}} &= \mathcal{C}^{T}\Big(\underbrace{\nabla_{\eta}{\tilde{p}}}_{=\mathbf{1}}\circ\underbrace{\nabla_{\tilde{p}}{f_{\tilde{p}}}}_{=\nabla_{\eta}{f_{\tilde{p}}}}\Big) = \mathcal{C}^{T}\nabla_{\eta}{f_{\tilde{p}}}\in{\mathbb{R}^{m}}\,.
\end{align}
The latter expression is efficiently evaluable, since $\mathcal{C}$ can be precomputed before optimization, i.e., $\nabla_{\phi}{f_{\tilde{p}}}$ only depends on $\nabla_{\eta}{f_{\tilde{p}}}$. It also follows that the reformulation-based optimization problem is convex if $f_{\tilde{p}}$ is convex in $\tilde{p}$ and the model-free subproblem is convex too.

\subsubsection{Optimization functions}
\begin{figure}[htb]
	\begin{center}
		\begin{tikzpicture}[node distance=3.7cm]
            
    \node[circle,draw,minimum size=1.4cm] at (0,0) (Phi) {$\phi$};
    \node[circle,draw,right of=Phi,yshift=1cm,minimum size=1.4cm] (D) {$d$};
    \node[circle,draw,right of=Phi,yshift=-1cm,minimum size=1.4cm] (DLETd) {$d\circ{l_{d}}$};
    \node[circle,draw,right of=Phi,xshift=1.5cm,yshift=-2.5cm,minimum size=0.7cm] (BRF) {$b$};
    \node[circle,draw,right of=D,yshift=-1cm,minimum size=1.4cm] (Eta) {$\eta$};
    \node[circle,draw,right of=Eta,xshift=-1.5cm,yshift=2cm,minimum size=0.7cm] (K) {$k$};
    \node[circle,draw,right of=Eta,yshift=0cm,minimum size=1.4cm] (P) {$p$};
    \node[circle,draw,right of=P,yshift=0cm,minimum size=1.4cm] (fP) {$f_{p/\tilde{p}}$};
    
    \draw[-latex] (Phi) -- (D);
    \draw[-latex] (Phi) -- (DLETd);
    \draw[-latex] (D) -- (Eta);
    \draw[-latex] (DLETd) -- (Eta);
    \draw[-latex] (BRF) -- (Eta);
    \draw[-latex] (K) -- (P);
    \draw[-latex] (Eta) -- (P);
    \draw[-latex] (P) -- (fP);

    \draw[-latex,violet,dashed] (fP) edge[bend right,draw=none] coordinate[at start](fP-b) coordinate[at end](P-b) (P)
    edge[bend left,draw=none] coordinate[at start](fP-t) coordinate[at end](P-t) (P)
    (fP-t) to [bend right=0] node [below, sloped] (fPGrad) {$\nabla_{p}{f_{p}}$} (P-t);
    \draw[-latex,violet,dashed] (P) edge[bend right,draw=none] coordinate[at start](P-b) coordinate[at end](Eta-b) (Eta)
    edge[bend left,draw=none] coordinate[at start](P-t) coordinate[at end](Eta-t) (Eta)
    (P-t) to [bend right=0] node [below, sloped] (PGrad) {$\nabla_{\eta}{p}$} (Eta-t);
    \draw[-latex,violet,dashed] (Eta) edge[bend right,draw=none] coordinate[at start](Eta-b) coordinate[at end](DLETd-b) (DLETd)
    edge[bend left,draw=none] coordinate[at start](Eta-t) coordinate[at end](DLETd-t) (DLETd)
    (Eta-t) to [bend right=0] node [below, sloped] (EtaDLETdGrad) {$\beta_{2}$} (DLETd-t);
    \draw[-latex,violet,dashed] (Eta) edge[bend right,draw=none] coordinate[at start](Eta-b) coordinate[at end](D-b) (D)
    edge[bend left,draw=none] coordinate[at start](Eta-t) coordinate[at end](D-t) (D)
    (Eta-t) to [bend right=0] node [below, sloped] (EtaDGrad) {$\beta_{1}$} (D-t);
    \draw[-latex,violet,dashed] (DLETd) edge[bend right,draw=none] coordinate[at start](DLETd-b) coordinate[at end](Phi-b) (Phi)
    edge[bend left,draw=none] coordinate[at start](DLETd-t) coordinate[at end](Phi-t) (Phi)
    (DLETd-t) to [bend right=0] node [below, sloped] (DLETdGrad) {$\mathcal{M}$} (Phi-t);
    \draw[-latex,violet,dashed] (D) edge[bend right,draw=none] coordinate[at start](D-b) coordinate[at end](Phi-b) (Phi)
    edge[bend left,draw=none] coordinate[at start](D-t) coordinate[at end](Phi-t) (Phi)
    (D-t) to [bend right=0] node [below, sloped] (DGrad) {$\mathcal{D}$} (Phi-t);
    \draw [-latex,purple] (Eta) to [bend left=90]  node [below, sloped]  (SurrP) {$\tilde{p}$} (fP);
    \draw [-latex,purple,dashed] (fP) to [bend left=90]  node [below, sloped]  (ftPGrad) {$\nabla_{\eta}{f_{\tilde{p}}}$} (Eta);
\end{tikzpicture}
	\end{center}
    \vspace{-1em}
    \caption{\footnotesize\textbf{Forward and backward calculation tree for POLO model-based optimization functions.} The black solid arrows symbolize forward value passes, while the violet dashed arrows indicate backward gradient passes. Our linear reformulation bypasses the probability function $p$ and builds a direct path connection from $\eta$ to $f_{\tilde{p}}$ (purple arrows). We denote the fluence $\phi$, dose $d$, dose-averaged LET $l_{d}$, binary risk factor $b$, linear predictor $\eta$, volume correction factor $k$, POLO $p$, optimization function $f_{p/\tilde{p}}$, model coefficients $\beta_{1}$ and $\beta_{2}$, dose-influence matrix $\mathcal{D}$, and the \enquote{LET-times-dose-influence matrix} $\mathcal{M}$.}
    \label{fig:calc-tree}
\end{figure}
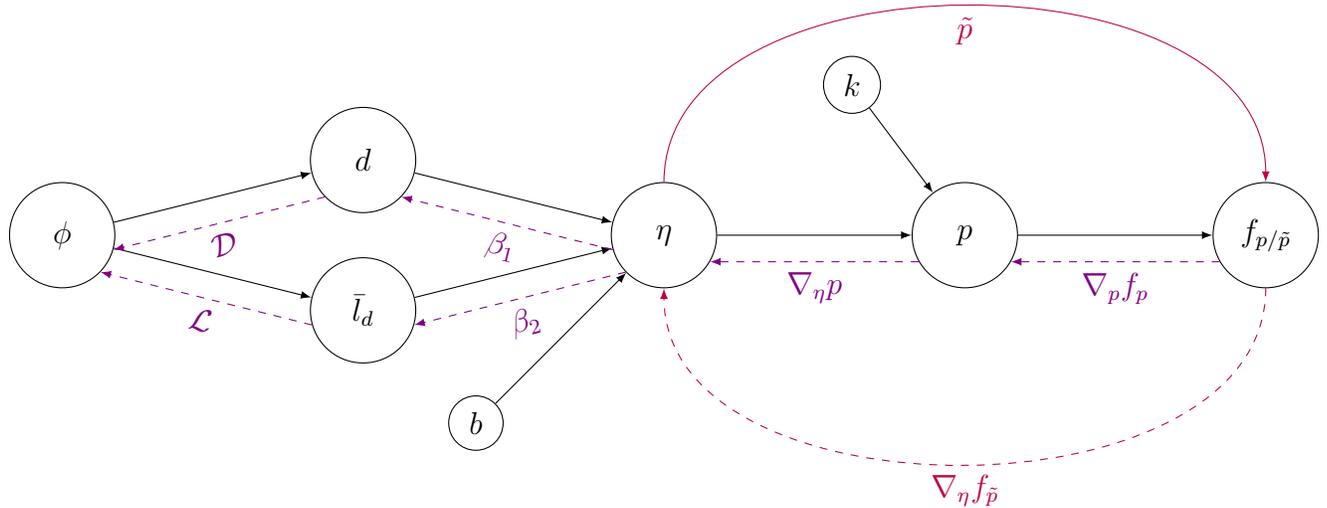

\Cref{fig:calc-tree} displays the corresponding forward and backward calculation tree. Given the importance of the choice for $f_{p/\tilde{p}}$, we propose feasible candidate functions in the following. For this purpose, we simplify the notation by omitting the arguments in the probability or decision function values, i.e., $p_{i}=p(\eta, k)_{i}$ and $\tilde{p}_{i}=\tilde{p}(\eta)_{i}$.

\paragraph{Seriality-assuming NTCP}
We can directly use the patient-level NTCP function from \Cref{eq:ntcp} as the optimization function $f_{p}$ with the probability function $p$. This function takes on a value close to zero if all $p_{i}$ are close to zero, and a value close to one if there is at least one $p_{i}$ close to 1. Differentiation by each $p_{i}$ yields the gradient function
\begin{align}\label{eq:ntcp-grad}
    \nabla_{p}{NTCP_{p}} = \begin{pmatrix} \prod_{j\neq{1}}^{n}{(1-p_{j})} & \cdots & \prod_{j\neq{n}}^{n}{(1-p_{j})} \end{pmatrix}^{T}\,.
\end{align}
Optimizing proton treatment plans for LGG patients with $NTCP_{p}$ has the advantage of acting directly on an interpretable outcome measure, with a degree of sensitivity to single high or many mid-range POLO values. 

\paragraph{Log-sum-exp approximation}
The probabilistic nature of $NTCP_{p}$ excludes the substitution of $\tilde{p}$ for $p$, i.e., the convexification of the optimization problem. To resolve this, we apply a smoothed maximum function via the logarithmic sum of exponentials (Log-sum-exp, LSE) approximation. The corresponding optimization function reads
\begin{align}
    LSE_{\tilde{p}} = \log{\left(\sum_{i=1}^{n}{\exp{(\tilde{p}_{i})}}\right)}\,,
\end{align}
where $\max{\{\tilde{p}_{1}, \ldots, \tilde{p}_{n}\}}$ could be added to account for numerical under-/overflow. $LSE_{\tilde{p}}$ is convex with gradient function
\begin{align}\label{eq:lse-grad}
    \nabla_{\tilde{p}}{LSE_{\tilde{p}}} = \begin{pmatrix} \frac{\exp{(\tilde{p}_{1})}}{\sum_{j=1}^{n}{\exp{(\tilde{p}_{j})}}} & \cdots & \frac{\exp{(\tilde{p}_{n})}}{\sum_{j=1}^{n}{\exp{(\tilde{p}_{j})}}} \end{pmatrix}^{T}\,.
\end{align}

\paragraph{Hellinger distance}
By interpreting the probability function values $p_{i}$ as realizations of random variables, one can also adopt statistical distance measures to design optimization functions $f_{p}$. The Hellinger distance is such an example, where the distance between the observed values $\begin{pmatrix} p_{1} & \cdots & p_{n}\end{pmatrix}^{T}$ and the reference values $\begin{pmatrix} q_{1} & \cdots & q_{n}\end{pmatrix}^{T}$ is measured by
\begin{align}\label{eq:hellinger-orig}
H_{p}=\frac{1}{\sqrt{2}}\sqrt{\sum_{i=1}^{n}{\left(\sqrt{p_{i}}-\sqrt{q_{i}}\right)^{2}}} \in{\left(0,\sqrt{\frac{n}{2}}\right)}\,.
\end{align}
Intuitively, the reference distribution for POLO model-based optimization should be $q_{i}=0$ for all $i\in\{1, \ldots, n\}$, i.e., should be the single-point zero distribution. By inserting $q_{i}=0$, removing the constant factor $\frac{1}{\sqrt{2}}$, and inverting the root function due to monotonicity, we obtain the transformed (and convex) optimization function
\begin{align}\label{eq:hellinger-opt}
	\tilde{H}_{p} := \sum_{i=1}^{n}{p_{i}} \in{(0, n)}\,,
\end{align}
which yields the constant gradient
\begin{align}
	\nabla_{p}{\tilde{H}_{p}} = \mathbf{1}\,.
\end{align}
We note that above gradient does not depend on the scaling of $p_{i}$, i.e., we could interchangeably use the decision function values $\tilde{p}_{i}$ in \Cref{eq:hellinger-opt}, even though the function $\tilde{H}_{\tilde{p}}$ would no longer be predicated on a probabilistic measure. The Hellinger distance can generally be thought of as a mean value reduction scheme, with less sensitivity on single high input function values than $NTCP_{p}$ and $LSE_{\tilde{p}}$.

\subsection{Implementation}\label{subsec:software}
To show the applicability of our approach towards treatment plan optimization, we implemented the concepts introduced in \cref{subsec:polo} and \cref{subsec:functions} in matRad (v2.10.1, \href{http://www.matRad.org}{http://www.matRad.org}) \cite{Wieser2017,Ackermann2020}, a MATLAB-based open-source software toolkit for radiotherapy treatment planning in education and research. This toolkit supports patient loading, dose/LET\textsubscript{d} calculation and optimization for intensity-modulated particle therapy, as well as plan evaluation and visualization.

For dose calculation we use matRad's pencil-beam algorithm with voxel raytracing \cite{Siddon1985,Hong1996} to compute the dose-influence matrix $\mathcal{D}$. Specifically, for protons, the voxel dose is determined as the product of a tabulated integrated depth-dependent property with a tabulated lateral distribution (forming the dose kernel), where the lateral distribution is convolved with the impinging lateral fluence distribution. The product between dose and dose-averaged LET is calculated according to \Cref{eq:dose-avg-let}, where matRad simplifies the computation by internally operating on the local product of LET and dose contributions (from the kernel database) and then directly mapping onto the quantity $(d\circ{l_{d}})$ via the fluence $\phi$.

Our implementation of the POLO model-based optimization framework includes new classes for model handling and the interface to the optimization algorithm. First, we added a \enquote{virtual} POLO segmentation over all voxels inside the whole brain and outside the gross tumor volume (GTV) and the \SI{4}{\milli\meter} ventricular fringe, as well as precomputed $b$ from \Cref{eq:brf}. Second, we declared the POLO model from \Cref{eq:orig-polo} and its linear reformulation from \Cref{eq:polo-surr} as subclasses of matRad's \texttt{BackProjection}, which facilitates iteration-wise forward calculation and backward differentiation up to the optimization function, according to \Cref{fig:calc-tree}, and inserted an index filter to calculate the POLO prediction values only in voxels where the physical dose is at least \SI{2.0}{\gray}. Third, we extended matRad's objective function catalogue with $NTCP_{p}$, $LSE_{\tilde{p}}$, $\tilde{H}_{p}$ and $\tilde{H}_{\tilde{p}}$. It is then sufficient to import the LGG test patient data, assign a POLO model-based optimization function to the POLO segment and start the treatment plan optimization process.

\subsection{Treatment planning}\label{subsec:design}
This section introduces the LGG sample patient case, the baseline treatment plan and the optimization scenarios covered later. For the remainder, the physical proton dose will be given in units of \si{\gray}, and the biologically-weighted proton dose in units of \si{\gray}(RBE). The displayed dose, dose-averaged LET and POLO distributions are consistently interpolated back to the treatment planning CT resolution ($0.6055\times{0.6055}\times{1.0}$ \si{\milli\meter\cubed}) for visualization. Volume contours are drawn thicker than the isolines referring to: \SI{50}{\percent}, \SI{90}{\percent}, \SI{100}{\percent} and \SI{110}{\percent} of the prescribed fractional dose to the CTV/PTV; 4, 5, 8 and 10 $\si{\keV\per\mu\meter}$ for the dose-averaged LET; and \SI{10}{\percent}, \SI{25}{\percent}, \SI{50}{\percent}, and \SI{80}{\percent} of the maximum observed POLO.

\subsubsection{Patient}
We selected a sample patient case from the cohort described by Sallem et al. \cite{Sallem2024}, who received proton therapy for LGG treatment in 2022. Particle radiation was delivered using three beams, with a total dose of \SI{54}{\gray}(RBE) at constant RBE of 1.1, applied in 30 fractions at the Heidelberg Ion Therapy Center (HIT), Germany. The clinical dose distribution was calculated using the Monte Carlo dose engine in RayStation 11B IonPG (RaySearch Laboratories, Stockholm, Sweden), configured with a dose uncertainty of \SI{0.5}{\percent} and a spatial dose resolution of $2.0\times{2.0}\times{2.0}$ \si{\milli\meter\cubed}. GTV and cerebral ventricles were manually delineated by a radiation oncologist on a T2-weighted MRI scan, while the remaining structures were contoured on the planning CT. In this case, the tumor volume was localized in the right temporo-insular region, with visible overlap between the PTV and the cerebral ventricles.

\subsubsection{Baseline plan}
We configured the baseline proton plan for the LGG sample patient in matRad with three couch angles of \ang{175}, \ang{240} and \ang{310} at fixed gantry angles. The number of fractions was set to 30. Further, we determined the spatial dose grid resolution to be $3.0\times{3.0}\times{3.0}$ \si{\milli\meter\cubed} and applied a constant RBE scheme of 1.1. For the optimization step, we compiled the clinical target functions from a template for the sample patient, generated by clinical experts, into a near-equivalent matRad treatment plan. \Cref{tab:objectives} summarizes the dose objectives, \Cref{tab:constraints} the dose constraints selected for the plan.

\begin{table}[htbp]
    \centering \renewcommand{\arraystretch}{1}
{\fontsize{9}{9} \selectfont
\setlength\tabcolsep{2pt}
\begin{tabular}{rccl}
        \textbf{Segment} & \textbf{Objective function} $f_{r}$ & \textbf{Objective weight} $w_{r}$ & \textbf{Dose parameters} \\ \toprule 
        Clinical Target Volume (CTV) & Squared Deviation & 100 & \SI{54}{\gray}(RBE) ref. \\
        & Dose Uniformity & 100 & -- \\
        Planning Target Volume (PTV) & Squared Deviation & 100 & \SI{54}{\gray}(RBE) ref. \\
        & Dose Uniformity & 100 & -- \\
        & Minimum DVH & 250 & $\geq$ \SI{97}{\percent} vol. at \SI{52}{\gray}(RBE) \\
        PTV-CTV & Squared Underdosing & 250 & \SI{52.5}{\gray}(RBE) min. \\
        & Maximum DVH & 250 & $\leq$ \SI{56}{\gray}(RBE) at \SI{0}{\percent} vol. \\
        \SI{3}{\milli\meter} Fringe in PTV & Maximum DVH & 25 & $\leq$ \SI{50.25}{\gray}(RBE) at \SI{0}{\percent} vol. \\
        Ventricular Fringe in PTV-GTV & Squared Underdosing & 150 & \SI{49}{\gray}(RBE) min. \\
        & Maximum DVH & 250 & $\leq$ \SI{52.3}{\gray}(RBE) at \SI{0}{\percent} vol. \\
        & Squared Deviation & 250 & \SI{51.3}{\gray}(RBE) ref. \\
        & Dose Uniformity & 250 & -- \\
        Cochlea (Right) & Maximum DVH & 100 & $\leq$ \SI{45}{\gray}(RBE) at \SI{0}{\percent} vol. \\
        Inner Ear (Right) & Maximum DVH & 150 & $\leq$ \SI{44}{\gray}(RBE) at \SI{0}{\percent} vol. \\\bottomrule
    \end{tabular}}

    \caption{\footnotesize\textbf{Dose objectives for the baseline plan.} The function names correspond to the dose objectives implemented in matRad. \hyperref[sec:matrad-functions]{\textbf{Appendix}} provides a mathematical definition of the objective functions used.}
    \label{tab:objectives}
\end{table}
\begin{table}[htbp]
    \centering \renewcommand{\arraystretch}{1.0}
{\fontsize{9}{9} \selectfont
\setlength\tabcolsep{6pt}
\begin{tabular}{rcl}
    \textbf{Segment} & \textbf{Constraint function} $c_{s}$ & \textbf{Dose parameters} $c_{s}^{l}$ / $c_{s}^{u}$ \\ \toprule
    Optical Nerve (Left) & Minimum/Maximum Dose & \SI{0}{\gray}(RBE) min. / \SI{52}{\gray}(RBE) max. \\
    Optical Nerve (Right) & Minimum/Maximum Dose & \SI{0}{\gray}(RBE) min. / \SI{52}{\gray}(RBE) max. \\
    Chiasm & Minimum/Maximum Dose & \SI{0}{\gray}(RBE) min. / \SI{52}{\gray}(RBE) max. \\
    Brainstem & Minimum/Maximum Dose & \SI{0}{\gray}(RBE) min. / \SI{53}{\gray}(RBE) max. \\
    External Contour & Minimum/Maximum Dose & \SI{0}{\gray}(RBE) min. / \SI{56}{\gray}(RBE) max. \\
    Planning Target Volume (PTV) & Minimum/Maximum Dose & \SI{0}{\gray}(RBE) min. / \SI{59.4}{\gray}(RBE) max. \\
    Optical System + \SI{2}{\milli\meter} & Minimum/Maximum Dose & \SI{0}{\gray}(RBE) min. / \SI{52}{\gray}(RBE) max. \\\bottomrule
\end{tabular}}
    \caption{\footnotesize\textbf{Dose constraints for the baseline plan.} The function names correspond to the dose constraints implemented in matRad. \hyperref[sec:matrad-functions]{\textbf{Appendix}} provides a mathematical definition of the constraint functions used.}
    \label{tab:constraints}
\end{table}

The underlying inverse planning problem is inherently multi-criteria, and we handle the nature of this problem by describing the total objective function $f$ as a weighted sum of the partial dose objectives $f_{r}$, i.e., the inverse planning problem reads
\begin{align}\label{eq:problem}
    \min_{\phi} f(\phi) = \sum_{r}{w_{r}f_{r}(\mathcal{D}_{r}\phi)}\,, \text{ s.t. } c_{s}^{l}\leq{c_{s}(\mathcal{D}_{s}\phi)}\leq{c_{s}^{u}}\quad\forall{s}\,,
\end{align}
where $w_{r}$ denotes the weight of the objective associated with the $r$-th segmented structure, $\mathcal{D}_{r}$ and $\mathcal{D}_{s}$ are the submatrices of the dose-influence matrix $\mathcal{D}$ row-filtered for the respective structure, and $c_{s}$ describes the constraint function for the $s$-th structure with the lower and upper bounds $c_{s}^{l}$ and $c_{s}^{u}$. To solve the inverse planning problem, we use the large-scale nonlinear interior-point quasi-Newton IPOPT solver with the \emph{ma57} algorithm \cite{Duff2004,Waechter2006}. 

\begin{figure}[htb]
    \begin{center}
        \setlength\tabcolsep{0pt} 
\begin{tabular}{@{} r M{0.333\linewidth} M{0.333\linewidth} M{0.333\linewidth} @{}}
  & $d_{RBE,fx}$\phantom{ABCD} & $l_{d}$\phantom{ABCD} & $p$\phantom{ABCD} \\
  & \includegraphics[width=1\linewidth, trim=4.7cm 10.2cm 2.9cm 9.9cm, clip=True]{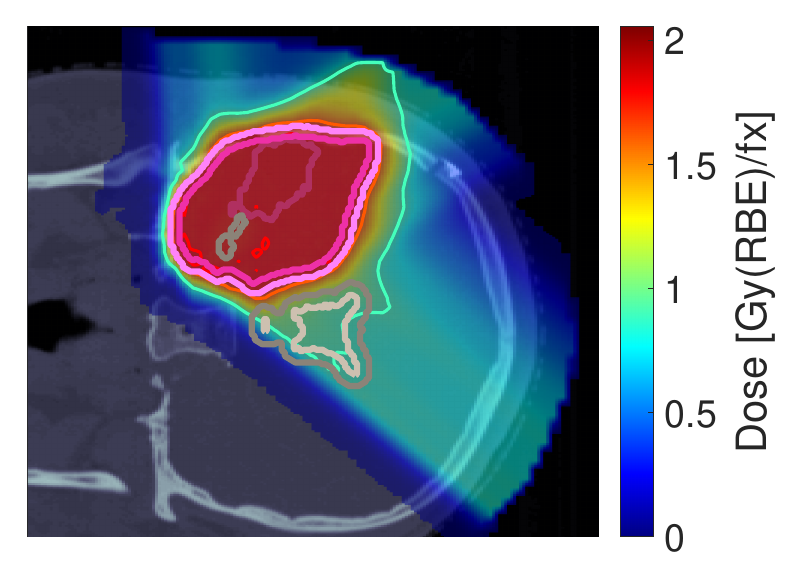} 
  & \includegraphics[width=1\linewidth, trim=4.7cm 10.2cm 2.9cm 9.9cm, clip=True]{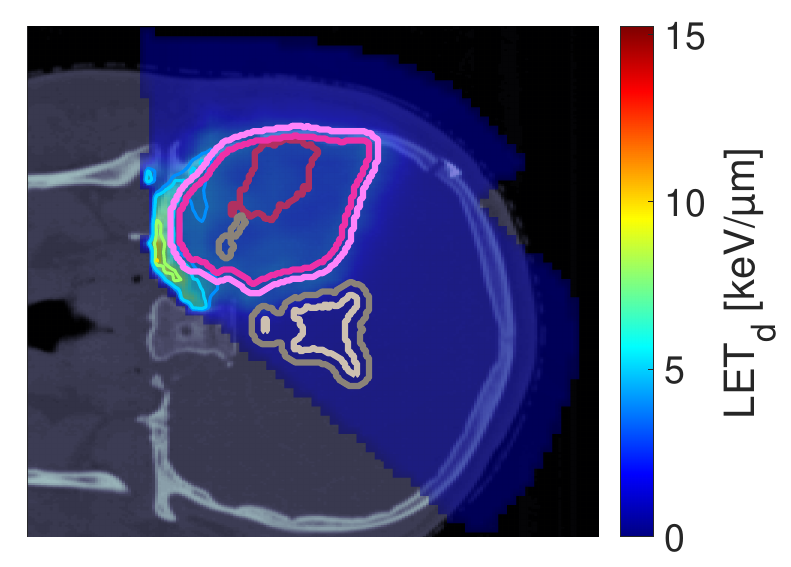} 
  & \includegraphics[width=1\linewidth, trim=4.7cm 10.2cm 2.9cm 9.9cm, clip=True]{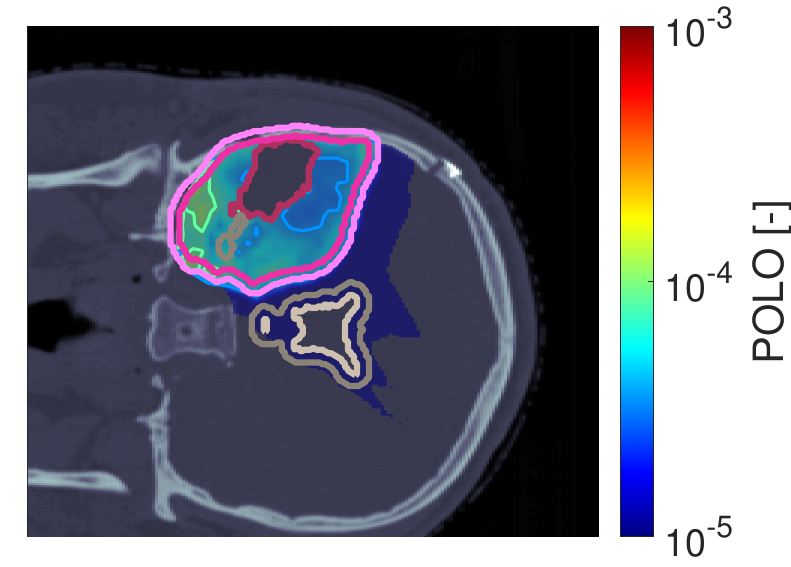} \\[-0.4em]
  & & & \text{\color{denim}{$NTCP:\SI{37.6}{\percent}$\phantom{ABCD}}} \\[-1.6em]
\end{tabular}
    \end{center}
    \caption{\footnotesize\textbf{Optimal 2D slice images of the RBE-weighted fractional dose $d_{RBE,fx}$, the dose-averaged linear energy transfer $l_{d}$ and the probability function values $p_{i}$ for the baseline plan.} The dose distribution (left) shows high target coverage with decreasing exposure beyond the CTV/PTV margins, but also visibly overlaps with parts of the VS. Meanwhile, for the dose-averaged LET (middle), layered homogeneity in the target volumes and local maxima at the distal ends of the proton beams can be observed. This results in more pronounced levels of POLO values within the target volumes, particularly in the presence of local dose-averaged LET maxima and in the environment of the VS. Referring back to the POLO model from \Cref{eq:orig-polo} and \eqref{eq:ext-polo}, we can expect exactly this type of coherence between the model output, the physical input feature profiles and the increased local risk near the VS. The baseline treatment plan exhibits a complication risk of \SI{37.6}{\percent} after evaluation of the POLO distribution with $NTCP_{p}$.}
    \label{fig:baseline-dose-let-polo}
\end{figure}

\Cref{fig:baseline-dose-let-polo} shows the optimal slice images of the dose (left), dose-averaged LET (middle) and POLO distribution (right). On the underlying CT scan, the tumor volumes are marked by violet to purple contours, while the VS (ventricles and \SI{4}{\milli\meter} fringe) is outlined in white and gray. We observe the high-dose region covering all of the GTV, with a largely homogeneous transition to the clinical and planning target volumes (CTV, PTV), and a moderate to severe overlap to the VS. The dose-averaged LET distributes in layers, with increased values around the contour edge of CTV/PTV and local peaks at the distal ends of the proton beams. Accordingly, hot spots on the POLO map are found in regions where dose and dose-averaged LET scale up, or close to the VS. Beyond the PTV, the POLO map reveals cold spots, due to the steepness of the dose and dose-averaged LET gradients.

\subsubsection{Optimization scenarios}
Our analysis on the feasibility and overall efficacy of POLO model-based treatment planning builds on a number of optimization scenarios. Here, we incorporated the POLO model-based optimization functions from \cref{subsec:functions} one by one as additional objectives in the weighted sum term from \Cref{eq:problem} while retaining the structure of the baseline proton plan, and manually tuned the weights associated with the POLO model-based objectives to generate treatment plans at near-equivalent NTCP levels of $29.7-30.5$ \si{\percent}, $19.6-20.3$ \si{\percent} and $9.8-10.2$ \si{\percent} (calculated with $NTCP_{p}$). In total, this yielded 12 scenarios (4 objective functions, 3 risk levels), which allow to compare the optimal slice images and histograms of the RBE-weighted fractional dose $d_{RBE,fx}$, the dose-averaged linear energy transfer $l_{d}$ and the probability function values $p_{i}$, as well as to evaluate the outcome prediction graphs by iterating the POLO maps and feeding them to $NTCP_{p}$.


\section{Results} \label{sec:results}

\subsection{\texorpdfstring{Dose, LET\textsubscript{d} \& POLO distributions}{Dose, LETd and POLO distributions}} \label{subsec:slices}
In this section, we present the optimal slice images and histograms for the RBE-weighted fractional dose $d_{RBE,fx}$, the dose-averaged linear energy transfer $l_{d}$ and the probability of lesion origin $p$, as well as their interplay.

\subsubsection{\texorpdfstring{RBE-weighted fractional dose $d_{RBE,fx}$}{RBE-weighted fractional dose dRBEfx}}
\begin{figure}[htbp]
    \begin{center}
        \setlength\tabcolsep{0pt} 
\centering
\begin{tabular}{@{} r M{0.315\linewidth} M{0.315\linewidth} M{0.315\linewidth} @{}}
\footnotesize & \text{\color{myteal}{$NTCP:{29.7-30.5}$ \si{\percent}\hspace{8pt}}} & \text{\color{red}{$NTCP:{19.6-20.3}$ \si{\percent}\hspace{8pt}}} & \text{\color{orange}{$NTCP:{9.8-10.2}$ \si{\percent}\hspace{8pt}}}\\
  \rotatebox[origin=c]{90}{\hspace{16pt}$d_{RBE,fx}$ $(NTCP_{p})$}
  & \includegraphics[width=\linewidth, trim=4.7cm 9.9cm 4.1cm 9.9cm, clip=True]{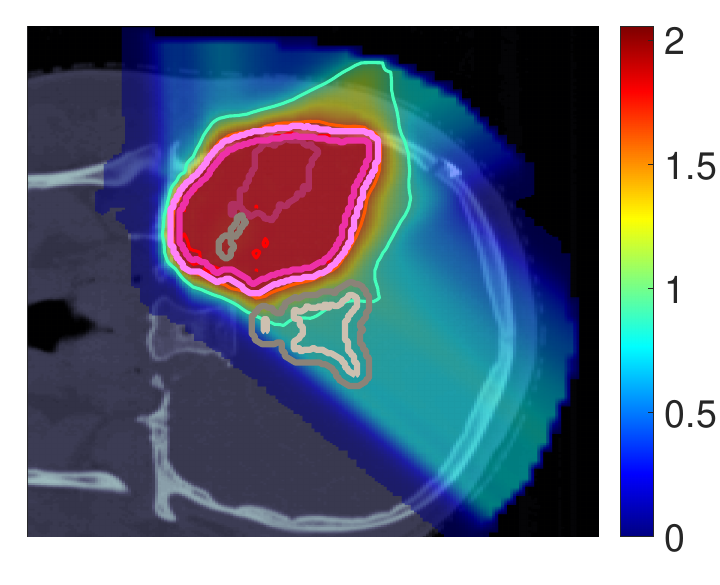} 
  & \includegraphics[width=\linewidth, trim=4.7cm 9.9cm 4.1cm 9.9cm, clip=True]{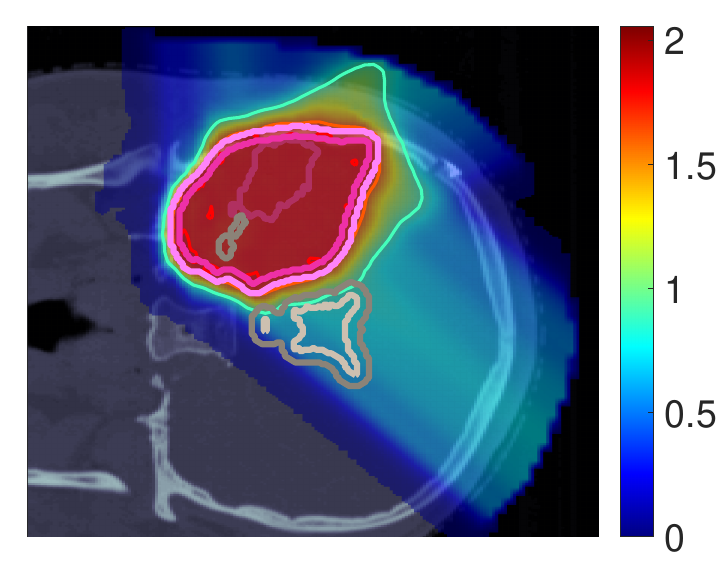}
  & \includegraphics[width=\linewidth, trim=4.7cm 9.9cm 4.1cm 9.9cm, clip=True]{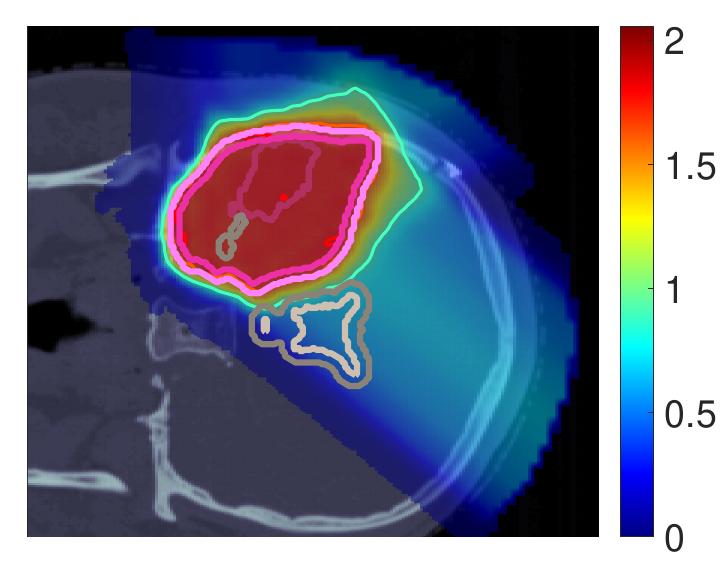} \\[-0.2em]
  \rotatebox[origin=c]{90}{\hspace{16pt}$d_{RBE,fx}$ $(LSE_{\tilde{p}})$}
  & \includegraphics[width=\linewidth, trim=4.7cm 9.9cm 4.1cm 9.9cm, clip=True]{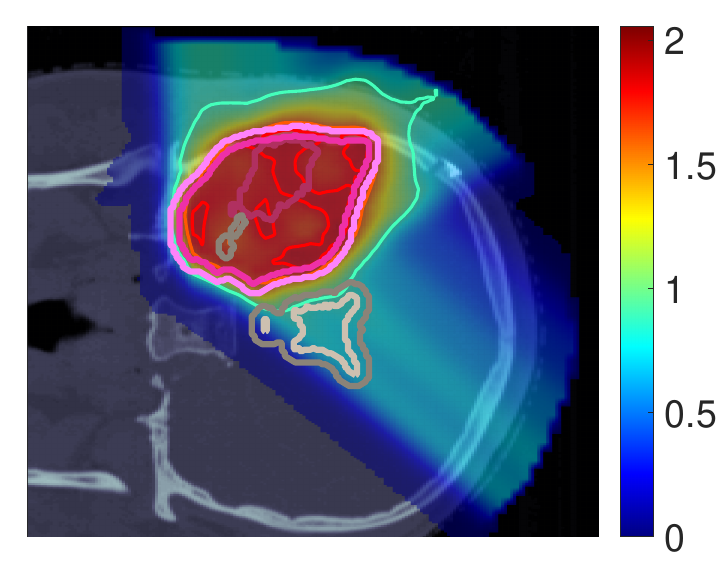} 
  & \includegraphics[width=\linewidth, trim=4.7cm 9.9cm 4.1cm 9.9cm, clip=True]{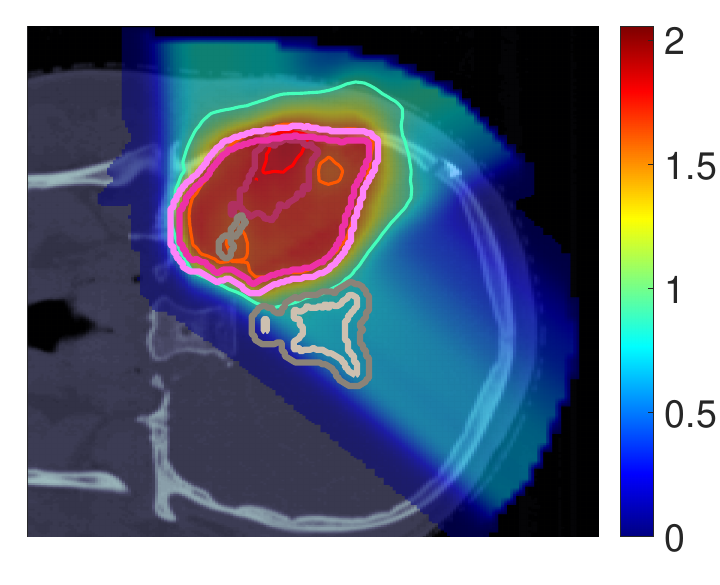}
  & \includegraphics[width=\linewidth, trim=4.7cm 9.9cm 4.1cm 9.9cm, clip=True]{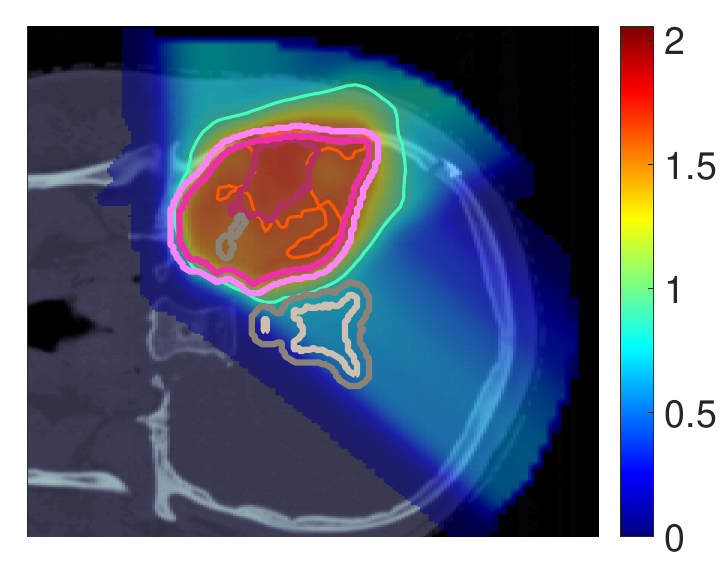} \\[-0.2em]
  \rotatebox[origin=c]{90}{\hspace{12pt}$d_{RBE,fx}$ $(\tilde{H}_{p})$}
  & \includegraphics[width=\linewidth, trim=4.7cm 9.9cm 4.1cm 9.9cm, clip=True]{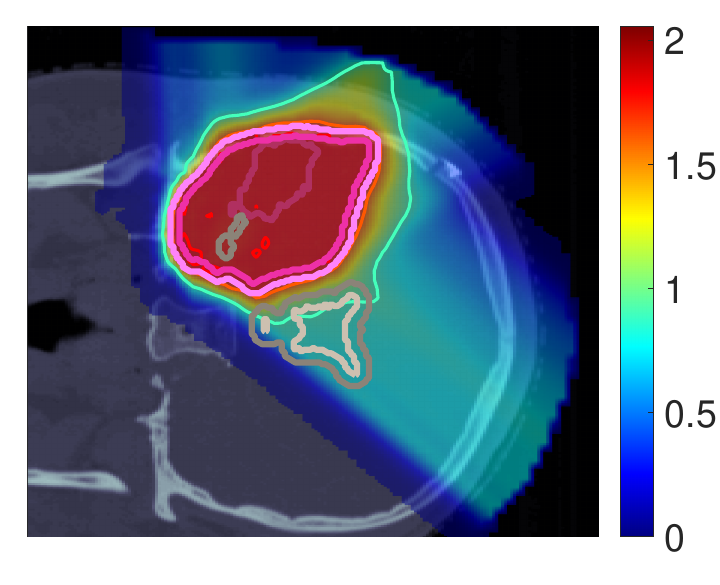} 
  & \includegraphics[width=\linewidth, trim=4.7cm 9.9cm 4.1cm 9.9cm, clip=True]{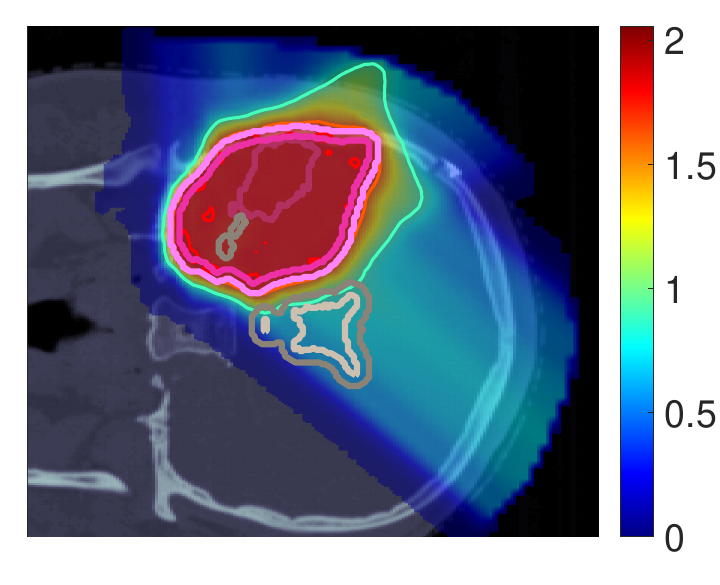}
  & \includegraphics[width=\linewidth, trim=4.7cm 9.9cm 4.1cm 9.9cm, clip=True]{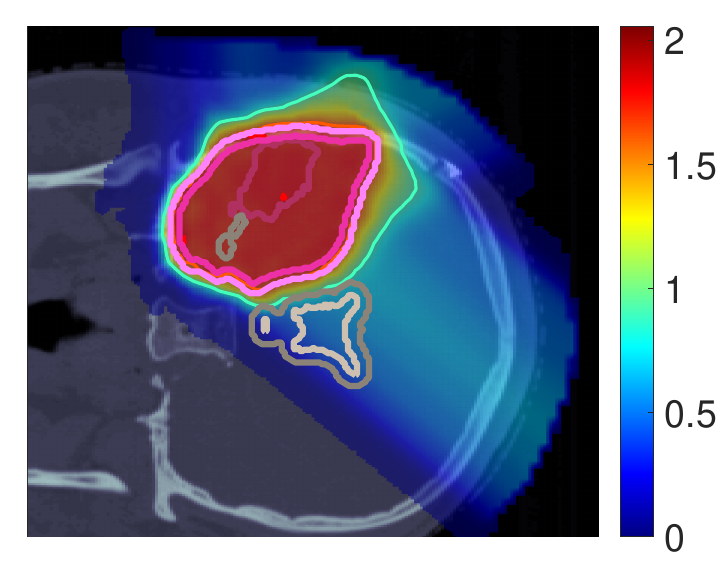} \\[-0.2em]
  \rotatebox[origin=c]{90}{\hspace{12pt}$d_{RBE,fx}$ $(\tilde{H}_{\tilde{p}})$}
  & \includegraphics[width=\linewidth, trim=4.7cm 9.9cm 4.1cm 9.9cm, clip=True]{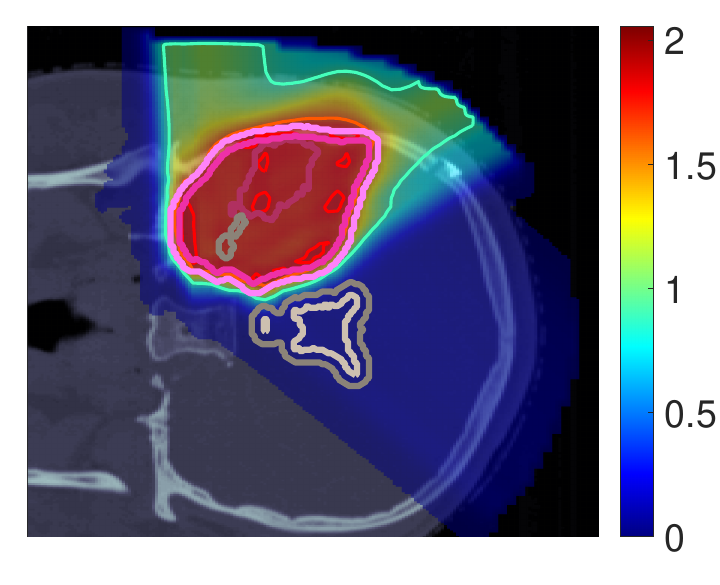} 
  & \includegraphics[width=\linewidth, trim=4.7cm 9.9cm 4.1cm 9.9cm, clip=True]{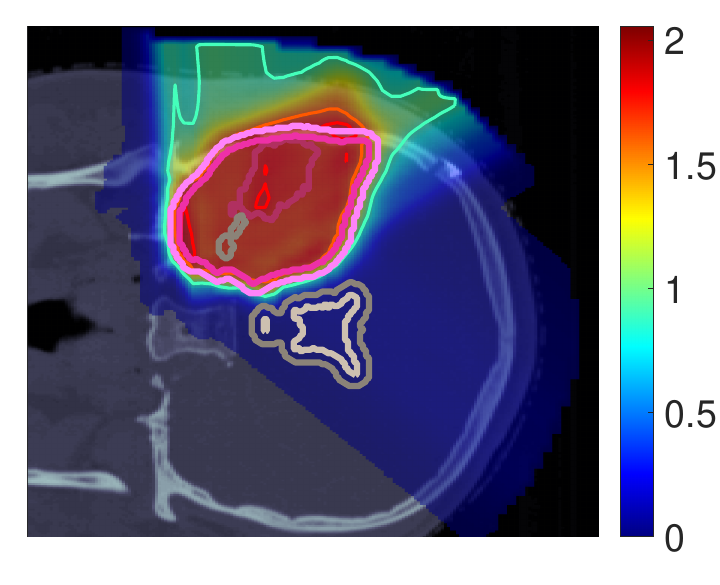}
  & \includegraphics[width=\linewidth, trim=4.7cm 9.9cm 4.1cm 9.9cm, clip=True]{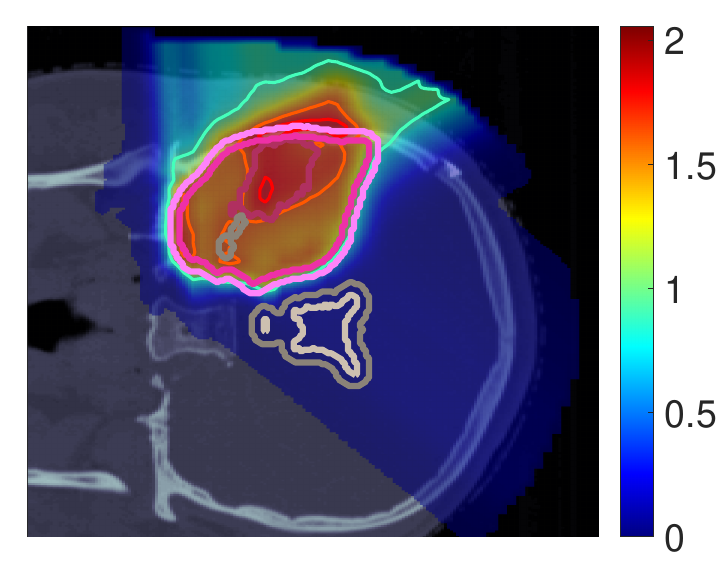} \\[-2em]
\end{tabular}
    \end{center}
\caption{\footnotesize\textbf{Optimal slice images of the RBE-weighted fractional dose $d_{RBE,fx}$ (in \si{\gray}(RBE)) for $NTCP_{p}$, $LSE_{\tilde{p}}$, $\tilde{H}_{p}$ and $\tilde{H}_{\tilde{p}}$ at different NTCP levels.} The objective functions behave similarly in that dose is reduced to achieve a lower NTCP value. However, they differ in the strength of the dose reduction inside and the impact on the dose outside the target volume, particularly in the VS. $NTCP_{p}$ and $\tilde{H}_{p}$ show comparable dose results, while $LSE_{\tilde{p}}$ and $\tilde{H}_{\tilde{p}}$ more aggressively eliminate dose in the target volume.}
\label{fig:optimal-dose}
\end{figure}

\Cref{fig:optimal-dose} shows the slice images with the optimal dose distributions, where the POLO model-based objective functions are displayed vertically and the NTCP levels horizontally. Throughout this part, we measure the loss in target coverage with respect to the baseline plan by the difference in RBE-weighted fractional doses received by \SI{95}{\percent} of the GTV and the PTV, denoted by $\Delta_{\text{GTV}} \text{D95}_{RBE,fx}$ and $\Delta_{\text{PTV}} \text{D95}_{RBE,fx}$ with units of \si{\gray}(RBE).

We observe intra-functional similarities of the objectives from a reduction of the dose values with decreasing NTCP level, especially in the overlapping regions to the VS, but also within the target volumes. $NTCP_{p}$ manages to reduce the risk estimate by almost \SI{18}{\pp} compared to the baseline plan without affecting target coverage ($\Delta_{\text{PTV}} \text{D95}_{RBE,fx}\approx{0.01}$, $\Delta_{\text{GTV}} \text{D95}_{RBE,fx}\approx{0.00}$), which can be attributed to the dose contraction away from the VS. This is even more pronounced when lowering the NTCP level to \SI{10}{\percent}, accompanied again by a preservation of target coverage ($\Delta_{\text{PTV}} \text{D95}_{RBE,fx}\approx{0.00}$, $\Delta_{\text{GTV}} \text{D95}_{RBE,fx}\approx{0.03}$).

From an inter-functional perspective $\tilde{H}_{p}$ seems to behave very similarly, with only minor deviations in the dose slice. In contrast, we find that $LSE_{\tilde{p}}$ and $\tilde{H}_{\tilde{p}}$ modulate the dose more strongly, with visible local inhomogeneities from the NTCP level of \SI{30}{\percent}. Here, $LSE_{\tilde{p}}$ only marginally reduces the dose in the lower part of the VS but restricts the dose exposure in the target volume ($\Delta_{\text{PTV}} \text{D95}_{RBE,fx}\approx{0.37}$, $\Delta_{\text{GTV}} \text{D95}_{RBE,fx}\approx{0.09}$). Even more, target coverage breaks down towards an NTCP level of \SI{10}{\percent} ($\Delta_{\text{PTV}} \text{D95}_{RBE,fx}\approx{0.6}$, $\Delta_{\text{GTV}} \text{D95}_{RBE,fx}\approx{0.23}$). One reason could be the focus of $LSE_{\tilde{p}}$ on maximum points, causing \enquote{notches} in the high dose region around the overlap of CTV/PTV and VS from the NTCP level of \SI{20}{\percent}. 

$\tilde{H}_{\tilde{p}}$, on the other hand, results in an immediate dose contraction, with higher concentration on the GTV with lower NTCP level. Similar to $LSE_{\tilde{p}}$, target coverage is drastically affected at \SI{10}{\percent} NTCP ($\Delta_{\text{PTV}} \text{D95}_{RBE,fx}\approx{0.72}$, $\Delta_{\text{GTV}} \text{D95}_{RBE,fx}\approx{0.19}$), which may be due to the increased scale of the function values and gradients of the linearly reformulated POLO model and therefore the higher impact of $LSE_{\tilde{p}}$ and $\tilde{H}_{\tilde{p}}$ on the plan. 

\begin{figure}[htbp]
    \begin{center}
        \setlength\tabcolsep{0pt} 
\centering
\begin{tabular}{@{} r M{0.5\linewidth} M{0.5\linewidth} @{}}
    & \includegraphics[width=0.95\linewidth, trim=7.8cm 7.35cm 8.8cm 7.3cm, clip=True]{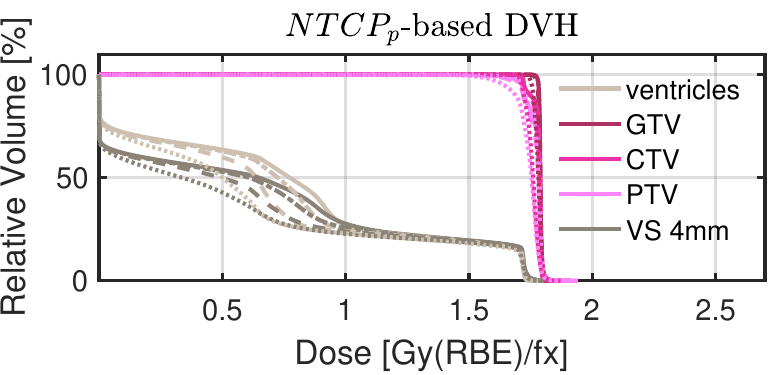}
    & \includegraphics[width=0.95\linewidth, trim=7.8cm 7.35cm 8.8cm 7.3cm, clip=True]{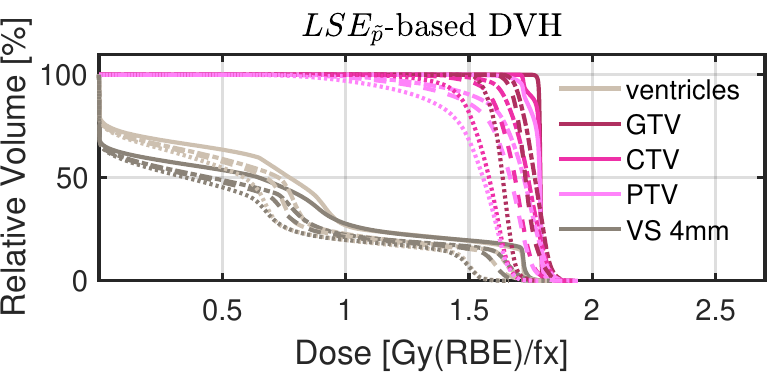} \\[-0.2em]
    & \includegraphics[width=0.95\linewidth, trim=7.8cm 7.35cm 8.8cm 7.3cm, clip=True]{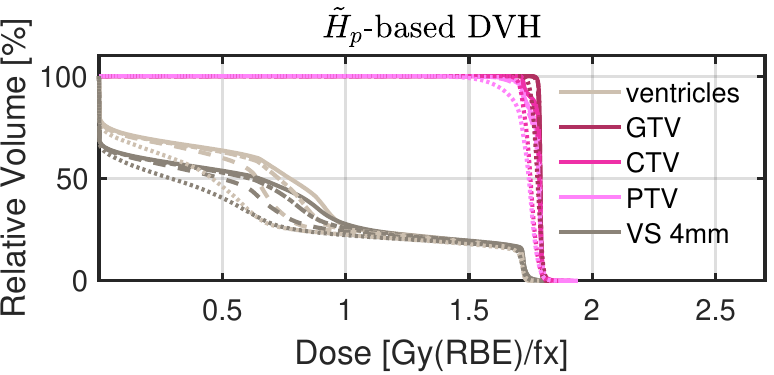}
    & \includegraphics[width=0.95\linewidth, trim=7.8cm 7.35cm 8.8cm 7.3cm, clip=True]{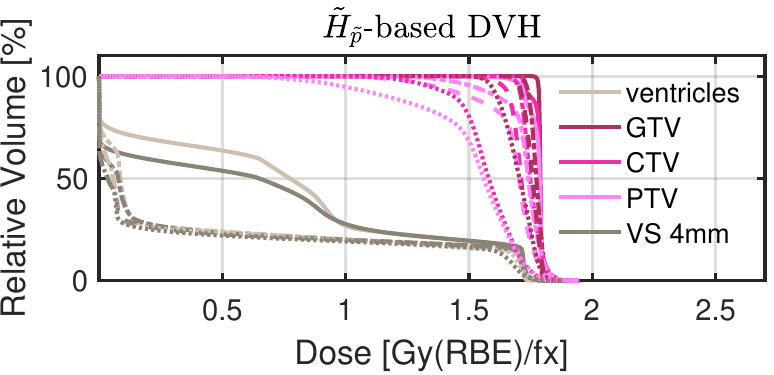} \\[-0.2em]
\end{tabular}
        \\[1em]
        \centering \renewcommand{\arraystretch}{1}
\setlength\tabcolsep{2pt}
{\fontsize{9}{9} \selectfont
  \begin{tabular}{l||C{1cm}C{1cm}C{1cm}C{1cm}||C{1cm}C{1cm}||C{1cm}C{1cm}||C{1cm}C{1cm}}
    \toprule
    \multirow{2}{*}{\tikz{\node[below left, inner sep=1pt] (def) {\textbf{Opt. case}\phantom{12}};%
      \node[above right,inner sep=1pt] (abc) {\phantom{12}\textbf{Segment}};%
      \draw (def.north west|-abc.north west) -- (def.south east-|abc.south east);}} &
      \multicolumn{2}{c}{ventricles} &
      \multicolumn{2}{c}{GTV} &
      \multicolumn{2}{c}{CTV} &
      \multicolumn{2}{c}{PTV} &
      \multicolumn{2}{c}{VS 4mm} \\
      & $D_{mean}$ & $D_{max}$ & $D5$ & $D95$ & $D5$ & $D95$ & $D5$ & $D95$ & $D_{mean}$ & $D_{max}$ \\
      \midrule
    Baseline ($NTCP=37.6\%$) \quad & 0.74 & 1.77 & 1.80 & 1.78 & 1.80 & 1.72 & 1.80 & 1.71 & 0.67 & 1.81 \\[1em]
    
    \quad\quad$+NTCP_{p}$ (30.2\%) & 0.72 & 1.77 & 1.80 & 1.78 & 1.80 & 1.72 & 1.80 & 1.71 & 0.64 & 1.81 \\
    \quad\quad$+NTCP_{p}$ (20.0\%) & 0.67 & 1.76 & 1.80 & 1.78 & 1.80 & 1.72 & 1.80 & 1.70 & 0.60 & 1.79 \\
    \quad\quad$+NTCP_{p}$ (10.2\%) & 0.61 & 1.78 & 1.80 & 1.75 & 1.79 & 1.71 & 1.79 & 1.71 & 0.56 & 1.78 \\[1em]
    
    \quad\quad$+LSE_{\tilde{p}}$ (30.5\%) & 0.66 & 1.84 & 1.84 & 1.69 & 1.83 & 1.61 & 1.83 & 1.34 & 0.58 & 1.84 \\
    \quad\quad$+LSE_{\tilde{p}}$ (20.3\%) & 0.63 & 1.81 & 1.82 & 1.66 & 1.77 & 1.54 & 1.77 & 1.25 & 0.56 & 1.81 \\
    \quad\quad$+LSE_{\tilde{p}}$ (9.8\%)\phantom{0} & 0.57 & 1.66 & 1.70 & 1.55 & 1.67 & 1.40 & 1.67 & 1.11 & 0.50 & 1.66 \\[1em]
    
    \quad\quad$+\tilde{H}_{p}$ (30.0\%) & 0.72 & 1.77 & 1.80 & 1.78 & 1.80 & 1.72 & 1.80 & 1.71 & 0.64 & 1.80 \\
    \quad\quad$+\tilde{H}_{p}$ (20.3\%) & 0.67 & 1.76 & 1.80 & 1.78 & 1.80 & 1.72 & 1.80 & 1.70 & 0.60 & 1.80 \\
    \quad\quad$+\tilde{H}_{p}$ (9.9\%)\phantom{0} & 0.59 & 1.79 & 1.80 & 1.74 & 1.79 & 1.71 & 1.79 & 1.62 & 0.54 & 1.79 \\[1em]
    
    \quad\quad$+\tilde{H}_{\tilde{p}}$ (29.7\%) & 0.41 & 1.80 & 1.81 & 1.73 & 1.82 & 1.70 & 1.82 & 1.51 & 0.40 & 1.82 \\
    \quad\quad$+\tilde{H}_{\tilde{p}}$ (19.6\%) & 0.39 & 1.84 & 1.80 & 1.70 & 1.79 & 1.64 & 1.79 & 1.36 & 0.38 & 1.84 \\
    \quad\quad$+\tilde{H}_{\tilde{p}}$ (10.2\%) & 0.38 & 1.88 & 1.79 & 1.59 & 1.75 & 1.34 & 1.75 & 0.99 & 0.36 & 1.88 \\
    \bottomrule
    \end{tabular}}
    \end{center}
    \vspace{-1em}
    \caption{\footnotesize\textbf{Optimal dose-volume histograms and dosimetrics at different NTCP levels for all POLO model-based objectives.} The histogram curves correspond to the target volumes and the VS, at different NTCP levels (~\SI{38}{\percent}: \protect\solid, ~\SI{30}{\percent}: \protect\dashdot., ~\SI{20}{\percent}: \protect\dashed, and ~\SI{10}{\percent}: \protect\dotted), and dosimetric values are reported with units of \si{\gray}(RBE) per fraction. We observe a positive correlation between dose values and NTCP for the VS in all cases. Additionally, for $NTCP_{p}$ and $\tilde{H}_{p}$, target coverage is not compromised, while for $LSE_{\tilde{p}}$ and $\tilde{H}_{\tilde{p}}$, the dose deposition in the target volumes decreases with NTCP.}
    \label{fig:dvh}
    \vspace{-10pt}
\end{figure}

Above results are underlined by \Cref{fig:dvh} which displays the corresponding dose-volume histograms (DVHs) and selected dosimetric values in each NTCP interval for the volumes of interest (VOIs) to allow for a global analysis of the dose distributions and their progression.

\subsubsection{\texorpdfstring{Dose-averaged linear energy transfer $l_{d}$}{Dose-averaged linear energy transfer ld}}
\begin{figure}[htbp]
    \begin{center}
        \setlength\tabcolsep{0pt} 
\centering

\begin{tabular}{@{} r M{0.315\linewidth} M{0.315\linewidth} M{0.315\linewidth} @{}}
\footnotesize & \text{\color{myteal}{$NTCP:{29.7-30.5}$ \si{\percent}\hspace{8pt}}} & \text{\color{red}{$NTCP:{19.6-20.3}$ \si{\percent}\hspace{8pt}}} & \text{\color{orange}{$NTCP:{9.8-10.2}$ \si{\percent}\hspace{8pt}}}\\
  \rotatebox[origin=c]{90}{\hspace{13pt}$l_{d}$ $(NTCP_{p})$}
  & \includegraphics[width=\linewidth, trim=4.7cm 9.9cm 4.1cm 9.9cm, clip=True]{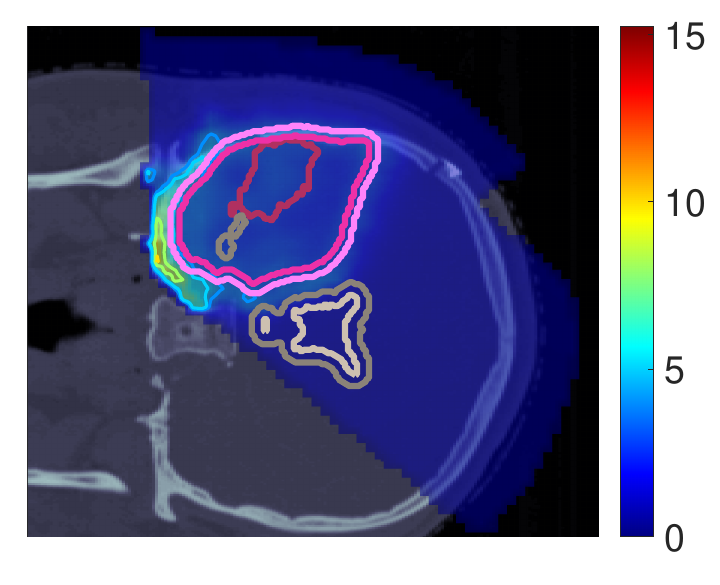} 
  & \includegraphics[width=\linewidth, trim=4.7cm 9.9cm 4.1cm 9.9cm, clip=True]{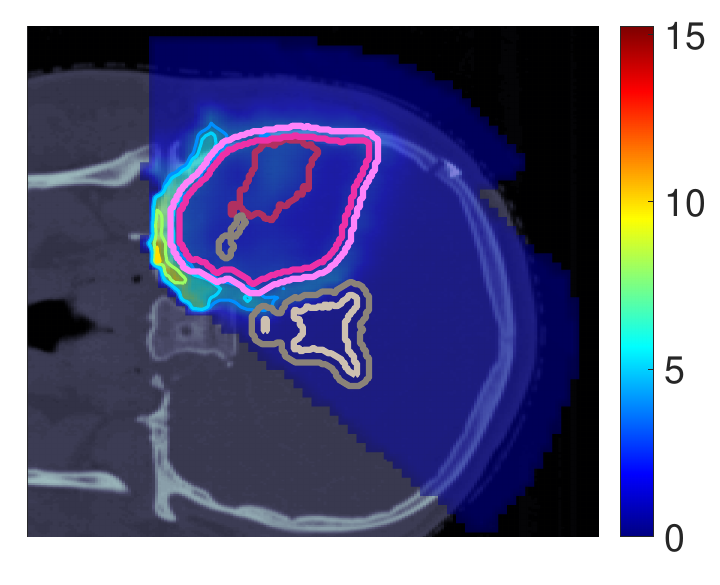}
  & \includegraphics[width=\linewidth, trim=4.7cm 9.9cm 4.1cm 9.9cm, clip=True]{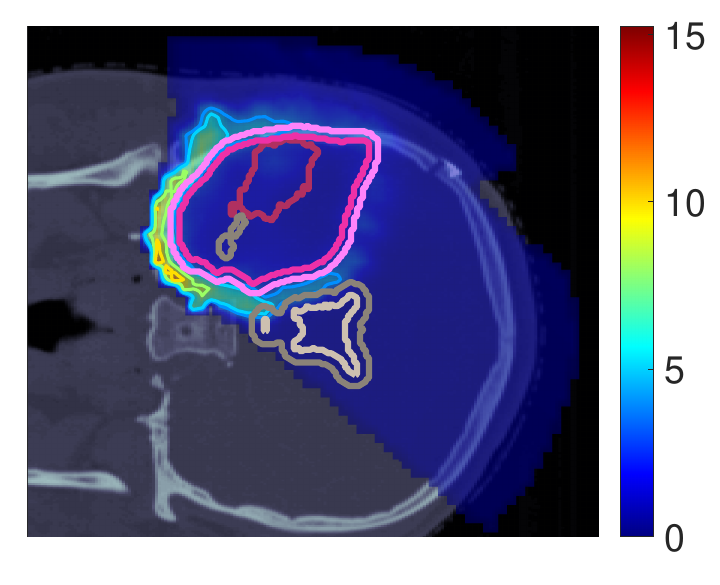} \\[-0.2em]
  \rotatebox[origin=c]{90}{\hspace{13pt}$l_{d}$ $(LSE_{\tilde{p}})$}
  & \includegraphics[width=\linewidth, trim=4.7cm 9.9cm 4.1cm 9.9cm, clip=True]{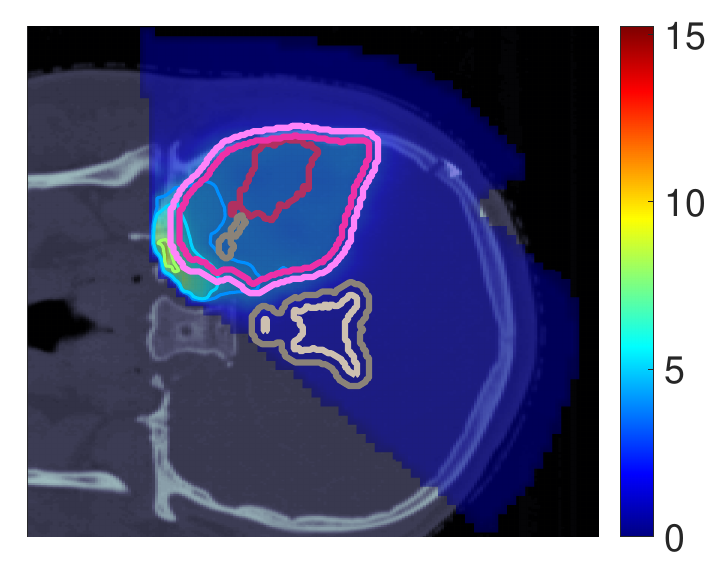} 
  & \includegraphics[width=\linewidth, trim=4.7cm 9.9cm 4.1cm 9.9cm, clip=True]{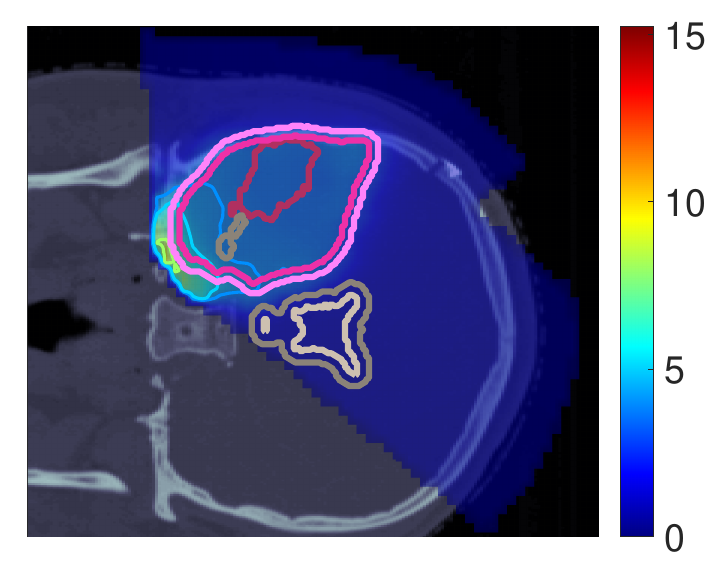}
  & \includegraphics[width=\linewidth, trim=4.7cm 9.9cm 4.1cm 9.9cm, clip=True]{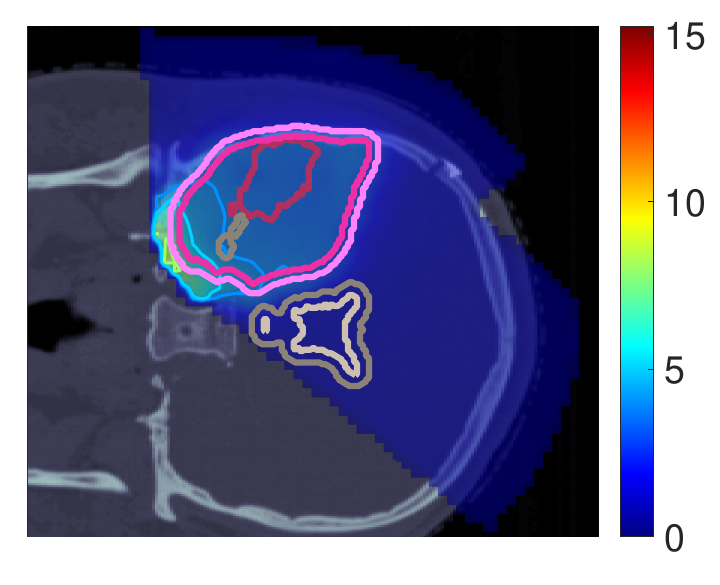} \\[-0.2em]
  \rotatebox[origin=c]{90}{\hspace{10pt}$l_{d}$ $(\tilde{H}_{p})$}
  & \includegraphics[width=\linewidth, trim=4.7cm 9.9cm 4.1cm 9.9cm, clip=True]{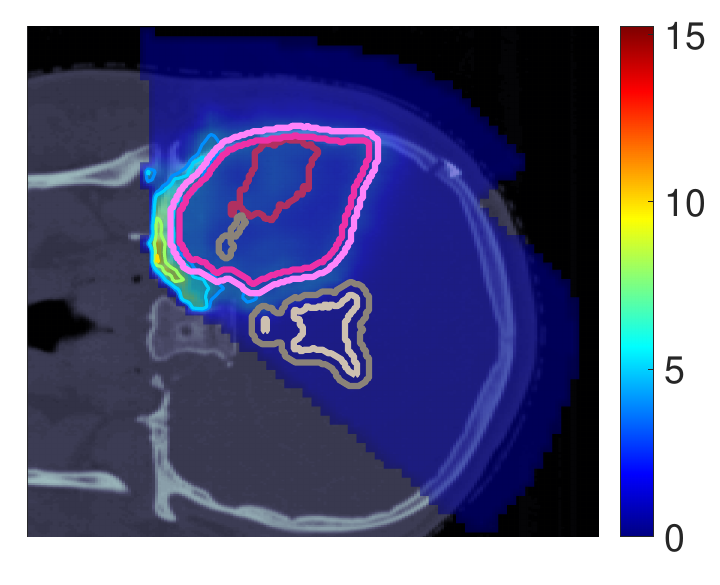} 
  & \includegraphics[width=\linewidth, trim=4.7cm 9.9cm 4.1cm 9.9cm, clip=True]{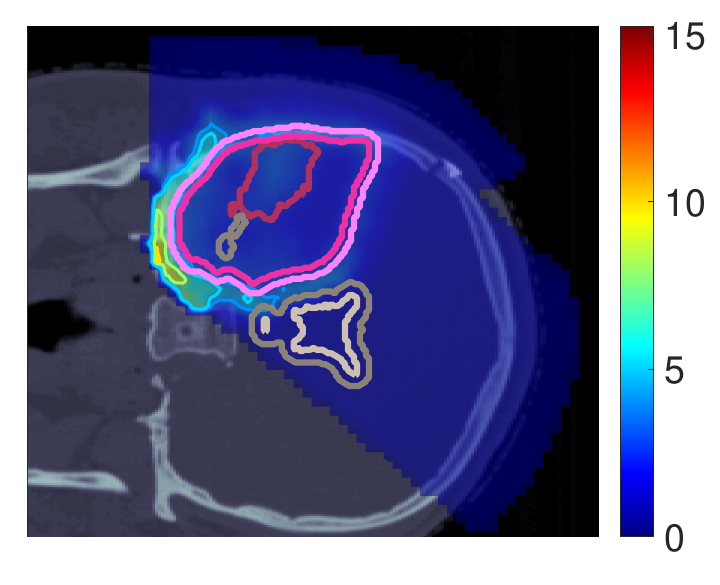}
  & \includegraphics[width=\linewidth, trim=4.7cm 9.9cm 4.1cm 9.9cm, clip=True]{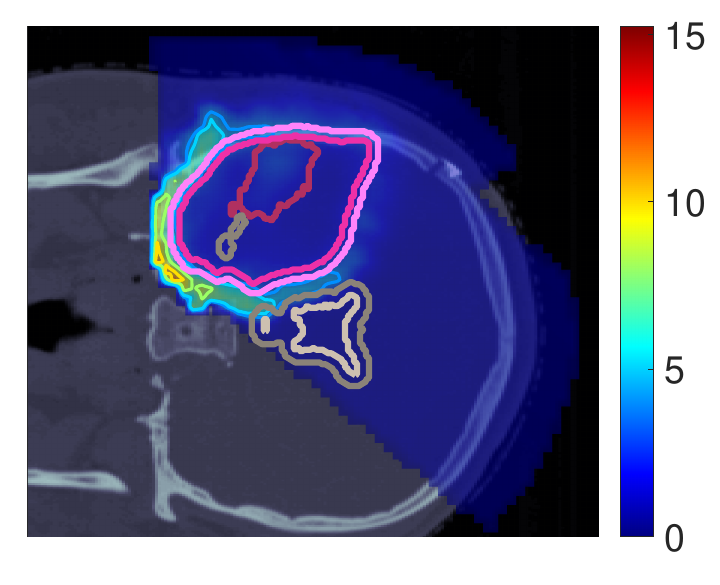} \\[-0.2em]
  \rotatebox[origin=c]{90}{\hspace{10pt}$l_{d}$ $(\tilde{H}_{\tilde{p}})$}
  & \includegraphics[width=\linewidth, trim=4.7cm 9.9cm 4.1cm 9.9cm, clip=True]{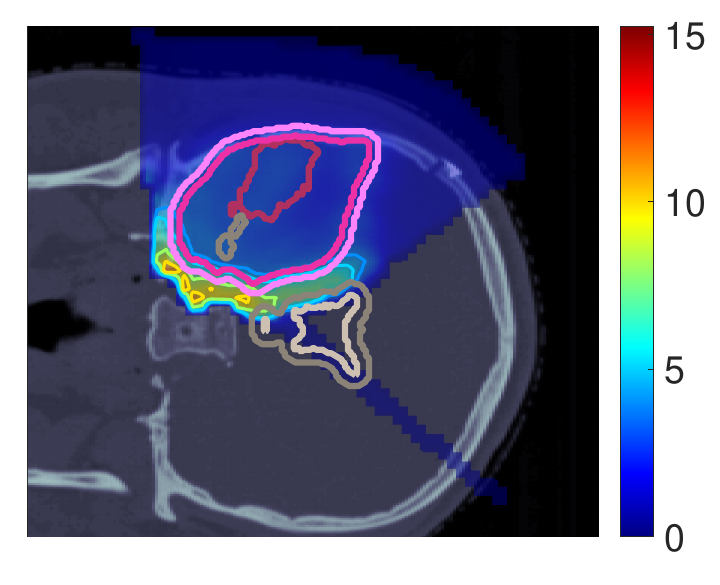} 
  & \includegraphics[width=\linewidth, trim=4.7cm 9.9cm 4.1cm 9.9cm, clip=True]{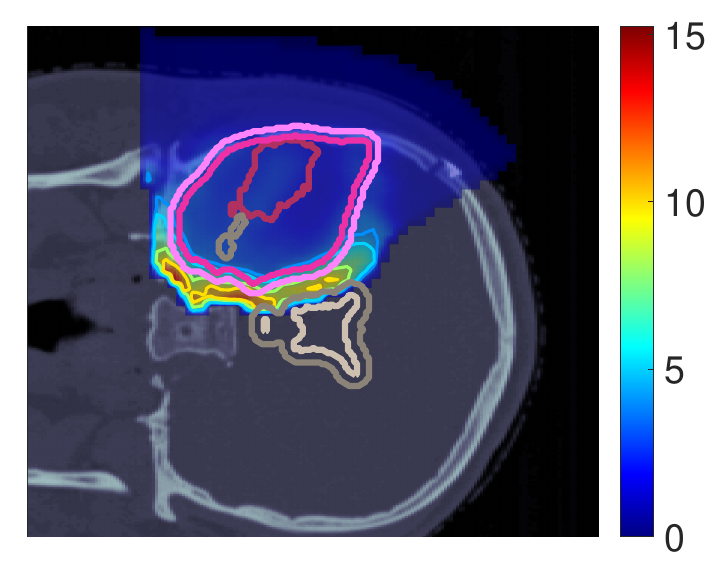}
  & \includegraphics[width=\linewidth, trim=4.7cm 9.9cm 4.1cm 9.9cm, clip=True]{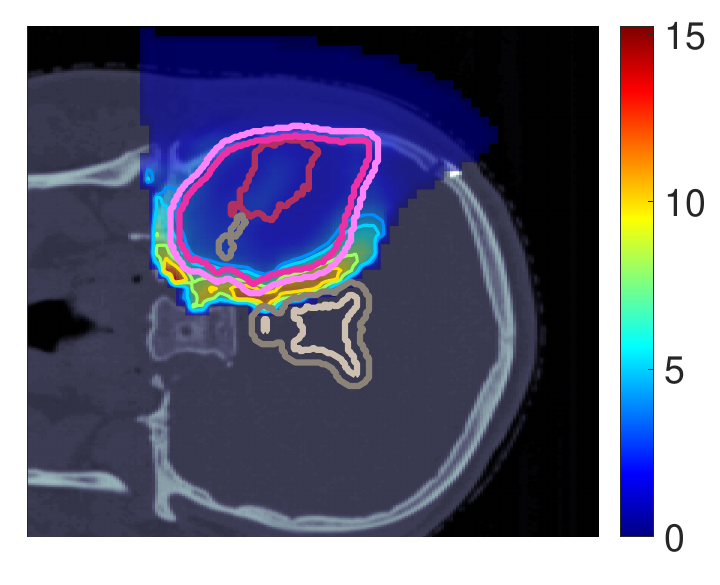} \\[-2em]
\end{tabular}
    \end{center}
\caption{\footnotesize\textbf{Optimal slice images of the dose-averaged linear energy transfer $l_{d}$ (in keV/$\boldsymbol{\mu}$m) for $NTCP_{p}$, $LSE_{\tilde{p}}$, $\tilde{H}_{p}$ and $\tilde{H}_{\tilde{p}}$ at different NTCP levels.} The shifts in $l_{d}$ coincide with those observed on the slice images for $d_{RBE,fx}$, i.e., there seems to be a trade-off between $d_{RBE,fx}$ and $l_{d}$ that is consistent with the mathematical structure of the POLO model. As a result, the $l_{d}$ slice images for $NTCP_{p}$ and $\tilde{H}_{p}$ are almost equal, while $LSE_{\tilde{p}}$ and $\tilde{H}_{\tilde{p}}$ differ. For the latter two objectives, the shift of $d_{RBE,fx}$ outside the target volume appears to be crucial, or, respectively, no shift of $l_{d}$ is necessary due to the strength of the reduction of $d_{RBE,fx}$.}
\label{fig:optimal-let}
\end{figure}

In light of the dose distribution changes and the interdependency between dose and dose-averaged LET, we can identify patterns in the slice images with the optimal dose-averaged LET distributions shown in \Cref{fig:optimal-let}. 

First, we note that $l_{d}$ hotspots do not manifest in the high dose regions -- this would be a risk driver because of the product term in the POLO model from \cref{subsec:polo}. Generally, some trade-off between $d_{RBE,fx}$ and $l_{d}$ is evident, where the optimizer allows for higher values of $l_{d}$ when $d_{RBE,fx}$ has been sufficiently reduced, and vice versa. 

For $NTCP_{p}$, the slight reduction of $d_{RBE,fx}$ outside the target volume coincides with a redistribution of $l_{d}$: higher values in the target volume are prevented, and instead induced along the margins of the PTV. Once again, $\tilde{H}_{p}$ behaves almost identically. Following the dose/LET\textsubscript{d} trade-off, the almost unchanged values for $LSE_{\tilde{p}}$ can also be explained: $LSE_{\tilde{p}}$ minimally reduces $d_{RBE,fx}$ in the region of overlap with the VS, meaning that high $l_{d}$ values there would be detrimental to the outcome prediction. Consequently, $l_{d}$ could only approach the \enquote{dose gap} that opens up within the target volume, or remain constant while the shifts in the dose distribution account for the reduction in NTCP. 

Finally, for $\tilde{H}_{\tilde{p}}$, we observe a sharp envelopment of the lower side of the PTV margin, consistent with the fall-off in $d_{RBE,fx}$ there. At an NTCP level of \SI{10}{\percent}, the high $l_{d}$ region has virtually manifested itself with hot spots partially overlapping the VS. However, revisiting the slice image for $d_{RBE,fx}$, this is not surprising, as in addition to the lower values in the target volume, the dissipation of $d_{RBE,fx}$ to the outside is only enhanced towards the upper end, leaving space for the $l_{d}$ band on the lower end, between the PTV margin and the VS.

\begin{figure}[htb]
    \begin{center}
        \setlength\tabcolsep{0pt} 
\centering
\begin{tabular}{@{} r M{0.5\linewidth} M{0.5\linewidth} @{}}
   & \includegraphics[width=0.95\linewidth, trim=7.8cm 7.35cm 8.8cm 7.3cm, clip=True]{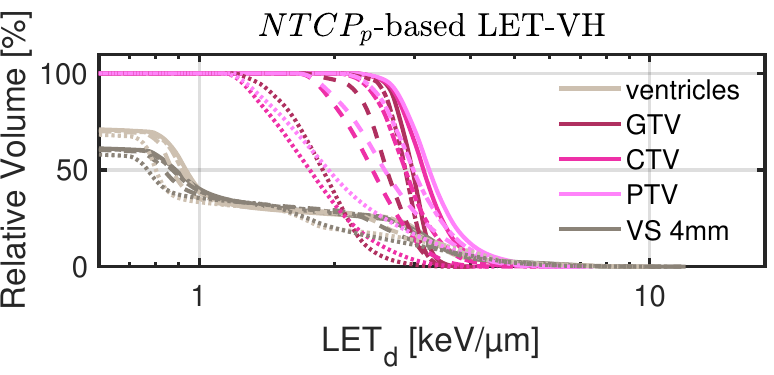}
   & \includegraphics[width=0.95\linewidth, trim=7.8cm 7.35cm 8.8cm 7.3cm, clip=True]{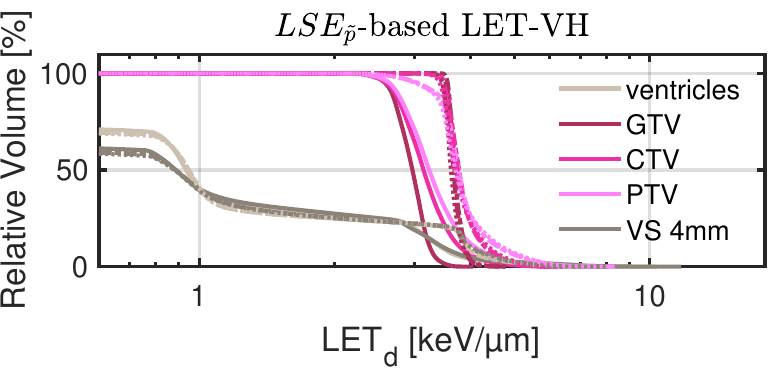} \\[-0.2em]
   & \includegraphics[width=0.95\linewidth, trim=7.8cm 7.35cm 8.8cm 7.3cm, clip=True]{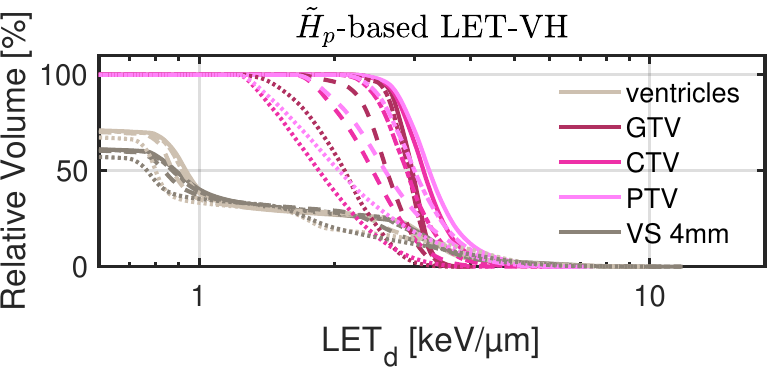}
   & \includegraphics[width=0.95\linewidth, trim=7.8cm 7.35cm 8.8cm 7.3cm, clip=True]{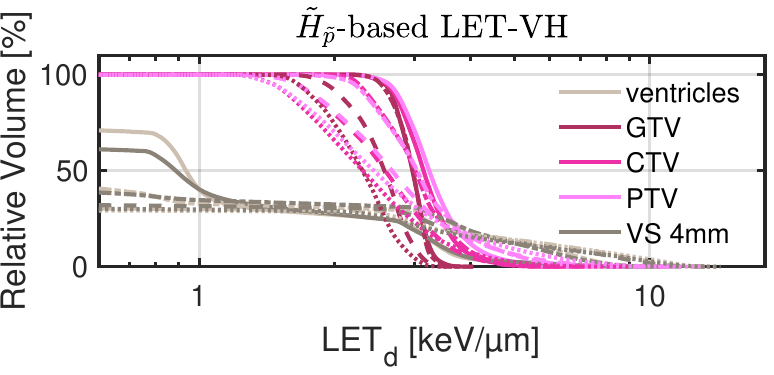} \\[-0.2em]
\end{tabular}
    \end{center}
    \vspace{-1em}
    \caption{\footnotesize\textbf{Optimal LET\textsubscript{d}-volume histograms at different NTCP levels for all POLO model-based objectives.} The histogram curves correspond to the target volumes and the VS, at different NTCP levels (~\SI{38}{\percent}: \protect\solid, ~\SI{30}{\percent}: \protect\dashdot., ~\SI{20}{\percent}: \protect\dashed, and ~\SI{10}{\percent}: \protect\dotted). We restrict the abscissa to values greater than or equal to 0.6, as the histogram curves are coherent below. With the exception of $LSE_{\tilde{p}}$, where only a slight increase in $l_{d}$ within the target volumes can be observed, all objectives reduce the $l_{d}$ values in the VOIs.}
    \label{fig:letvh}
\end{figure}

Again, we can confirm the slice images by the LET\textsubscript{d}-volume histograms (LET-VHs) from \Cref{fig:letvh}. 
For $NTCP_{p}$ and $\tilde{H}_{p}$, the values decrease with lower NTCP for all VOIs. $LSE_{\tilde{p}}$ almost does not change the distributions, only a slight increase at the left margin of the target volume is visible both in the slice images and in the respective LET-VH curves. Meanwhile, $\tilde{H}_{\tilde{p}}$ reduces the values in all VOIs, just as $NTCP_{p}$ and $\tilde{H}_{p}$, but in contrast, for the VS, the lower range is more attenuated and the upper range is enriched due to the redistribution along the PTV margin.

\subsubsection{\texorpdfstring{Probability of lesion origin $p$}{Probability of lesion origin POLO}}
\begin{figure}[htbp]
    \begin{center}
        \setlength\tabcolsep{0pt} 
\centering

\begin{tabular}{@{} r M{0.315\linewidth} M{0.315\linewidth} M{0.315\linewidth} @{}}
\footnotesize & \text{\color{myteal}{$NTCP:{29.7-30.5}$ \si{\percent}\hspace{8pt}}} & \text{\color{red}{$NTCP:{19.6-20.3}$ \si{\percent}\hspace{8pt}}} & \text{\color{orange}{$NTCP:{9.8-10.2}$ \si{\percent}\hspace{8pt}}}\\
  \rotatebox[origin=c]{90}{\hspace{13pt}$p$ $(NTCP_{p})$}\quad
  & \includegraphics[width=\linewidth, trim=4.7cm 9.9cm 3.85cm 9.9cm, clip=True]{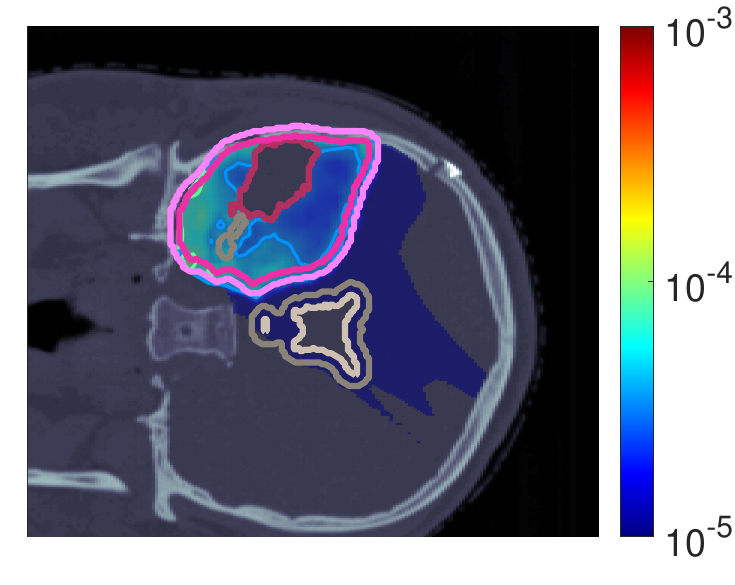} 
  & \includegraphics[width=\linewidth, trim=4.7cm 9.9cm 3.85cm 9.9cm, clip=True]{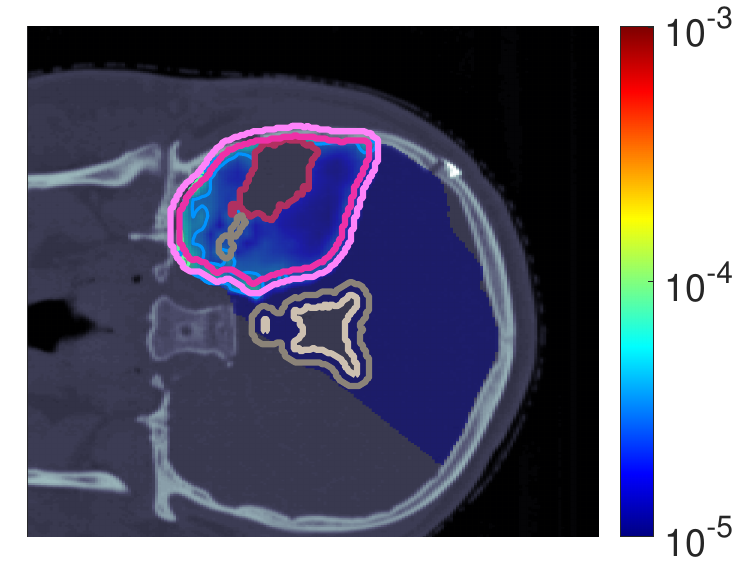}
  & \includegraphics[width=\linewidth, trim=4.7cm 9.9cm 3.85cm 9.9cm, clip=True]{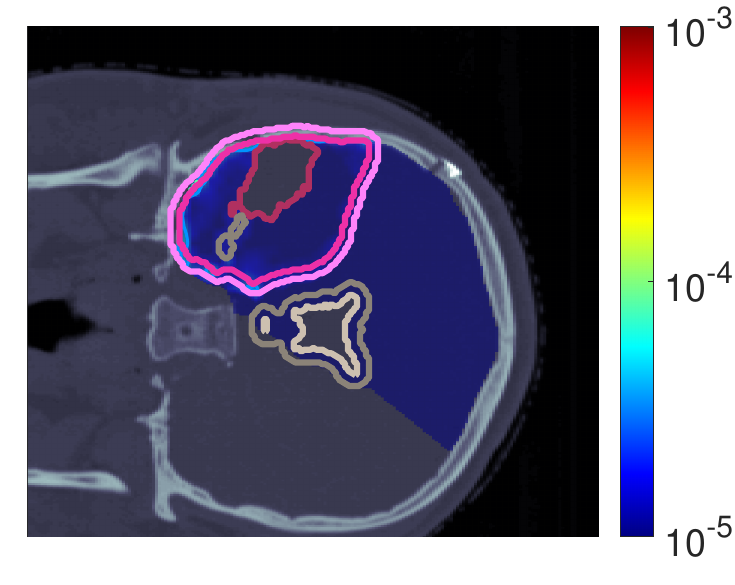} \\[-0.2em]
  \rotatebox[origin=c]{90}{\hspace{13pt}$p$ $(LSE_{\tilde{p}})$}\quad
  & \includegraphics[width=\linewidth, trim=4.7cm 9.9cm 3.85cm 9.9cm, clip=True]{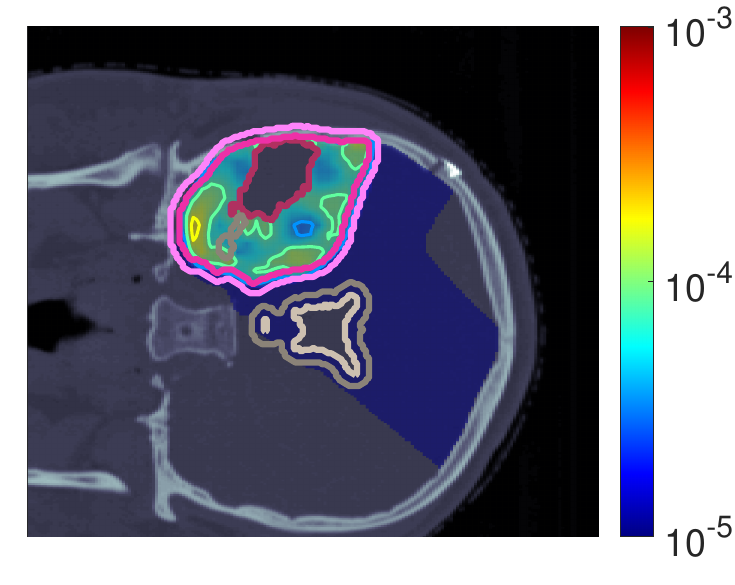} 
  & \includegraphics[width=\linewidth, trim=4.7cm 9.9cm 3.85cm 9.9cm, clip=True]{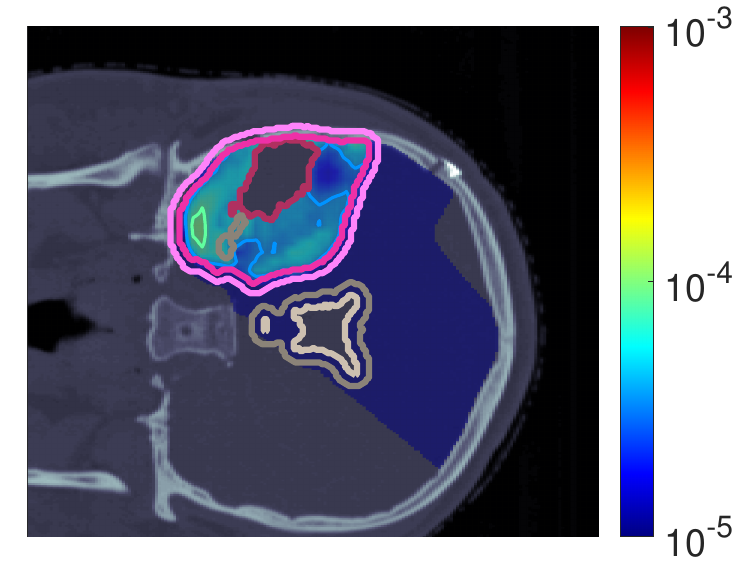}
  & \includegraphics[width=\linewidth, trim=4.7cm 9.9cm 3.85cm 9.9cm, clip=True]{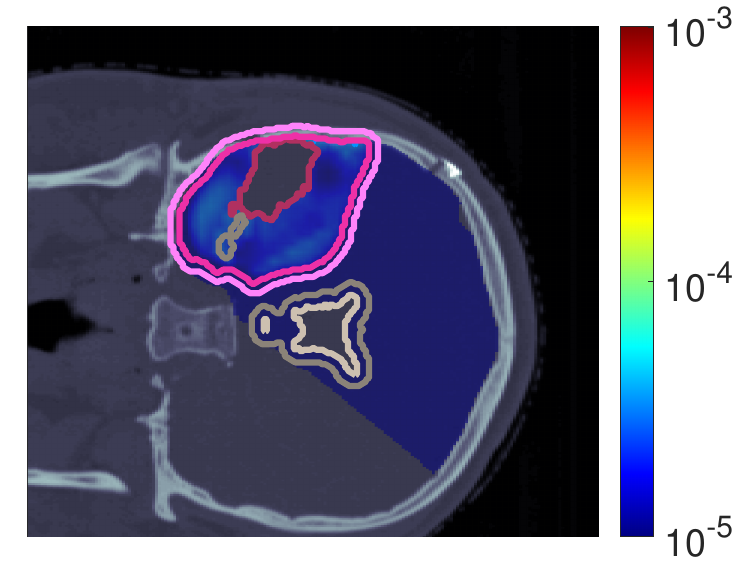} \\[-0.2em]
  \rotatebox[origin=c]{90}{\hspace{10pt}$p$ $(\tilde{H}_{p})$}\quad
  & \includegraphics[width=\linewidth, trim=4.7cm 9.9cm 3.85cm 9.9cm, clip=True]{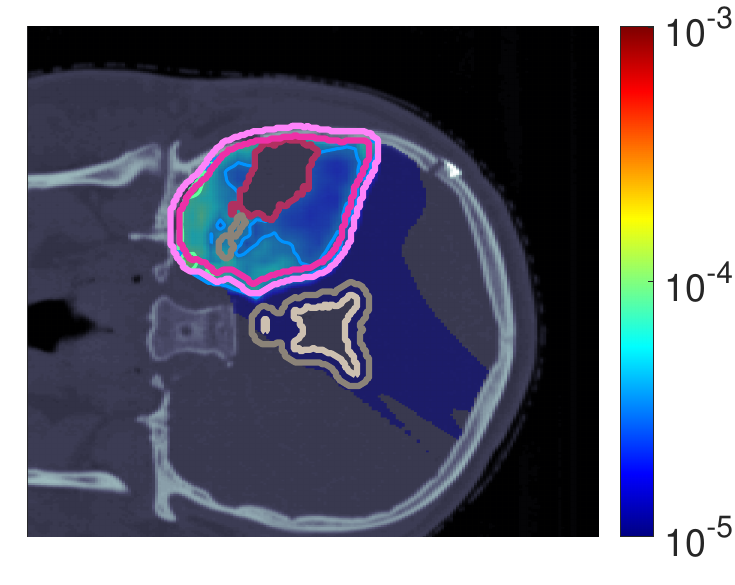} 
  & \includegraphics[width=\linewidth, trim=4.7cm 9.9cm 3.85cm 9.9cm, clip=True]{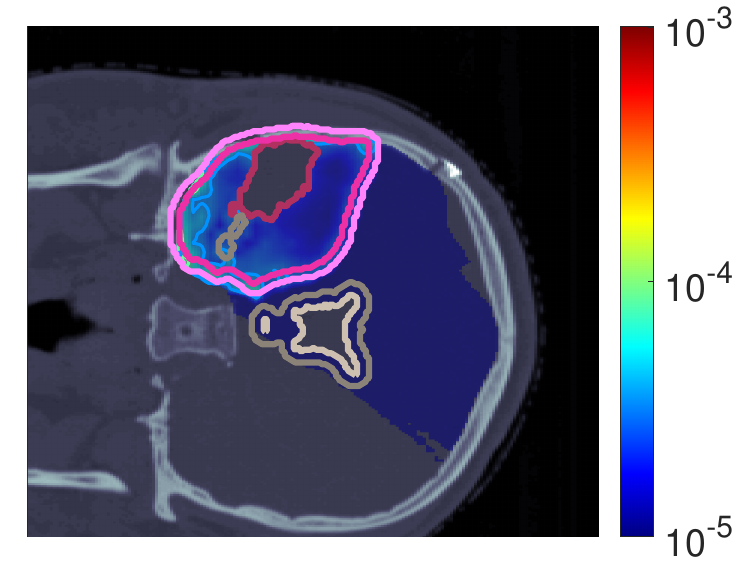}
  & \includegraphics[width=\linewidth, trim=4.7cm 9.9cm 3.85cm 9.9cm, clip=True]{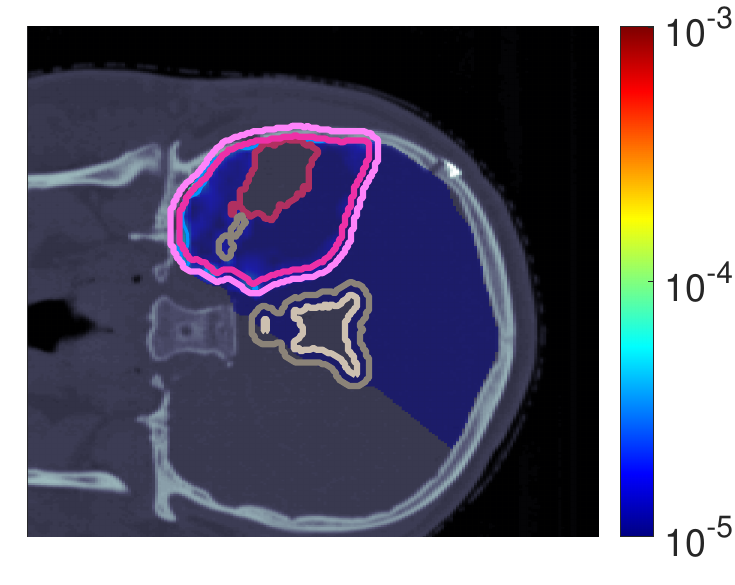} \\[-0.2em]
  \rotatebox[origin=c]{90}{\hspace{10pt}$p$ $(\tilde{H}_{\tilde{p}})$}\quad
  & \includegraphics[width=\linewidth, trim=4.7cm 9.9cm 3.85cm 9.9cm, clip=True]{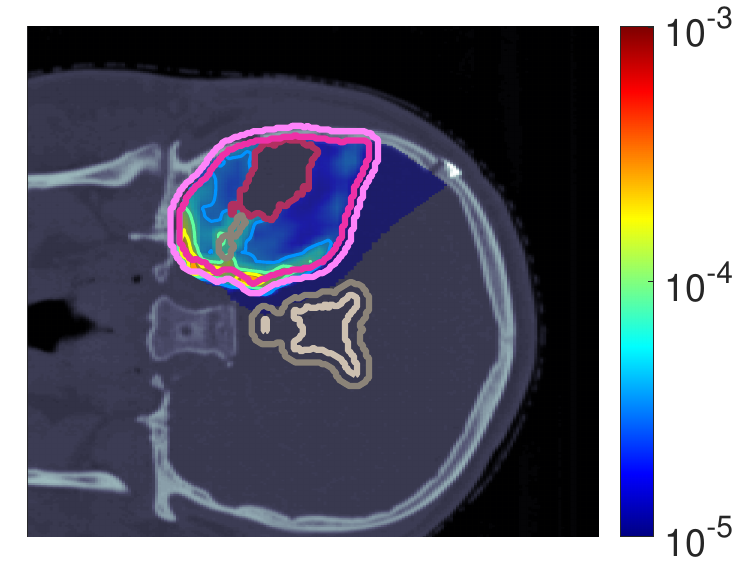} 
  & \includegraphics[width=\linewidth, trim=4.7cm 9.9cm 3.85cm 9.9cm, clip=True]{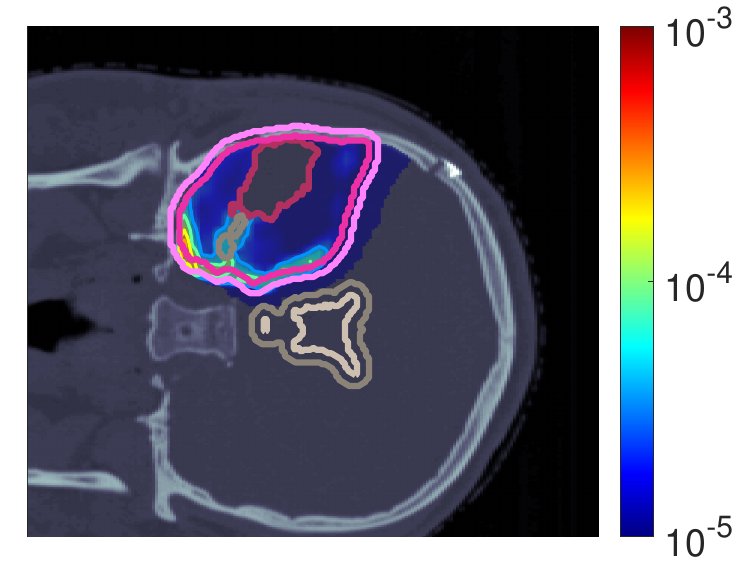}
  & \includegraphics[width=\linewidth, trim=4.7cm 9.9cm 3.85cm 9.9cm, clip=True]{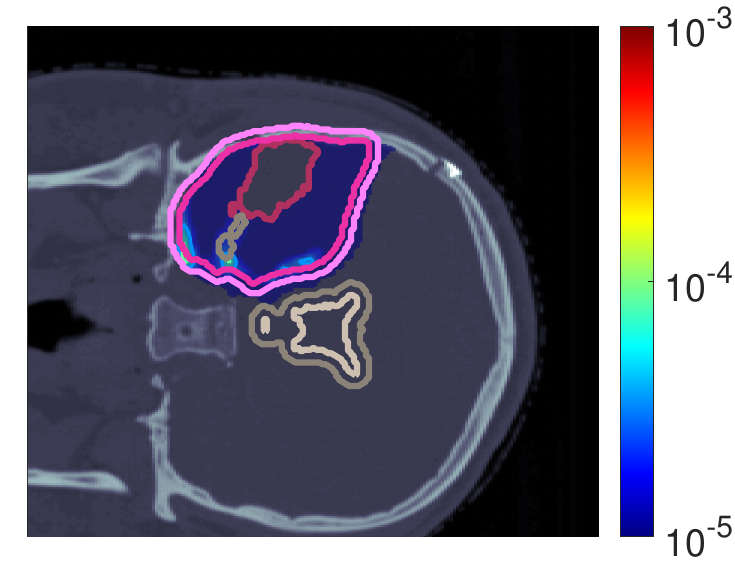} \\[-2em]
\end{tabular}
    \end{center}
\caption{\footnotesize\textbf{Optimal slice images of the probability of lesion origin $p$ for $NTCP_{p}$, $LSE_{\tilde{p}}$, $\tilde{H}_{p}$ and $\tilde{H}_{\tilde{p}}$ at different NTCP levels.} Considering the slice images for $d_{RBE,fx}$ and $l_{d}$ from \Cref{fig:optimal-dose} and \Cref{fig:optimal-let}, the calculated $p$ values can be deduced from the prediction model. For example, hot spots are primarily found where high $d_{RBE,fx}$ values meet high $l_{d}$ values, and for the voxels inside the \SI{4}{\milli\meter} ventricular fringe where $b$ boosts the probability prediction. We take from these results that the POLO model-based optimization of the LGG sample patient's proton plan achieves its goal, i.e., reducing $p$ in the region of interest.}
\label{fig:optimal-polo}
\end{figure}

\Cref{fig:optimal-polo} shows the POLO maps resulting from voxel-wise evaluation of the sigmoid-transformed linear predictor $\eta$ from \Cref{eq:orig-polo} with the input slice images for $d_{RBE,fx}$ and $l_{d}$.

For $NTCP_{p}$ and $\tilde{H}_{p}$, we observed a (slight) reduction of $d_{RBE,fx}$ in the target volume and in the lower region overlapping with the VS, as well as a redistribution of $l_{d}$ from the target volume to the margin of the PTV. Correspondingly, we see an overall reduction of $p$ in the target volume, together with an increasing attenuation of the hot spots around the marginal region of the PTV, and an elimination at the lower end of the VS. At strong down-regulation of the NTCP, the values are one order of magnitude smaller, with peaks in the intersections to the \SI{4}{\milli\meter} ventricular fringe and in regions where $d_{RBE,fx}$ and $l_{d}$ intensify. 

$LSE_{\tilde{p}}$ showed a decay of $d_{RBE,fx}$ inside but not outside the target volume, and almost constant values of $l_{d}$, and we recognize this pattern again in the $p$ distributions: hot spots are softened by the isolation of high-dose and high-LET\textsubscript{d} regions, and the $p$ values in the target volume follow the negative trend for $d_{RBE,fx}$ at lower NTCP levels. Even the \enquote{dose gap} that occurs at an NTCP of \SI{20}{\percent} in the lower and upper parts of the target volume can be read from $p$. Last, for $\tilde{H}_{\tilde{p}}$ we can correlate the results for $d_{RBE,fx}$ and $l_{d}$ again to understand $p$. The higher $l_{d}$ along the PTV margin is reflected by larger $p$ values and local hot spots, while the region around the GTV exhibits smaller $p$ values due to the reduction of $d_{RBE,fx}$. At the lowest NTCP level, $d_{RBE,fx}$ contracts around the GTV, leaving only a slightly upward region at the left margin of the PTV which resembles the high-LET\textsubscript{d} region, and hot spots where $d_{RBE,fx}$, $l_{d}$ and $b$ interact reinforcingly.

\begin{figure}[htb]
    \begin{center}
        \setlength\tabcolsep{0pt} 
\centering
\begin{tabular}{@{} r M{0.5\linewidth} M{0.5\linewidth} @{}}
    & \includegraphics[width=0.95\linewidth, trim=7.6cm 7cm 8.8cm 7.1cm, clip=True]{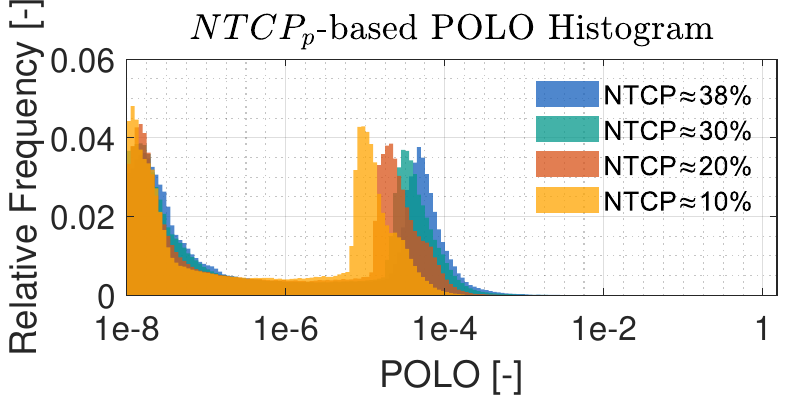}
    & \includegraphics[width=0.95\linewidth, trim=7.6cm 7cm 8.8cm 7.1cm, clip=True]{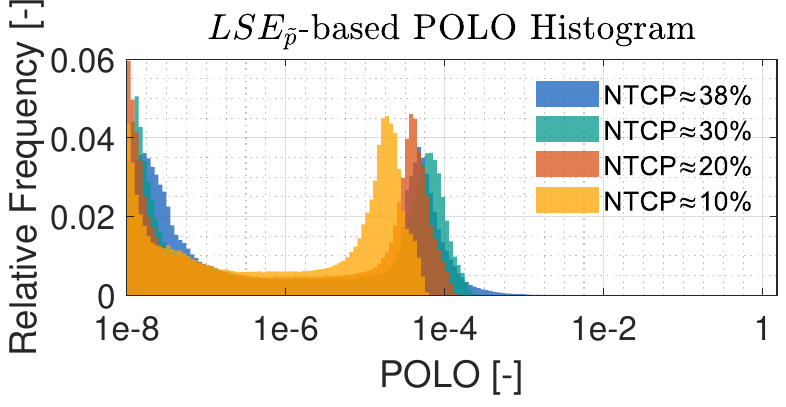} \\[-0.2em]
    & \includegraphics[width=0.95\linewidth, trim=7.6cm 7cm 8.8cm 7.1cm, clip=True]{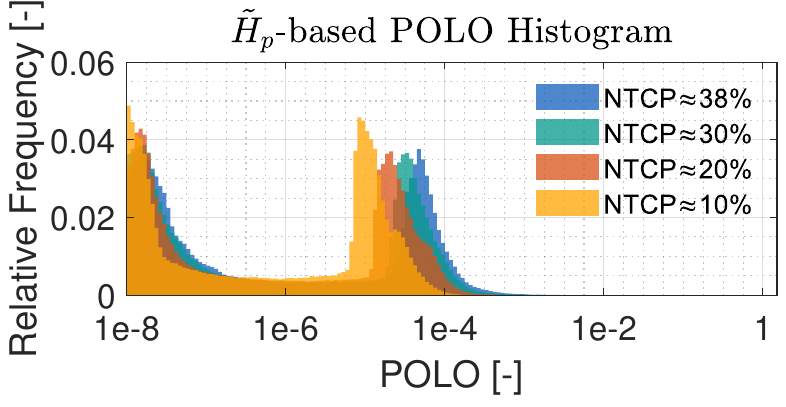}
    & \includegraphics[width=0.95\linewidth, trim=7.6cm 7cm 8.8cm 7.1cm, clip=True]{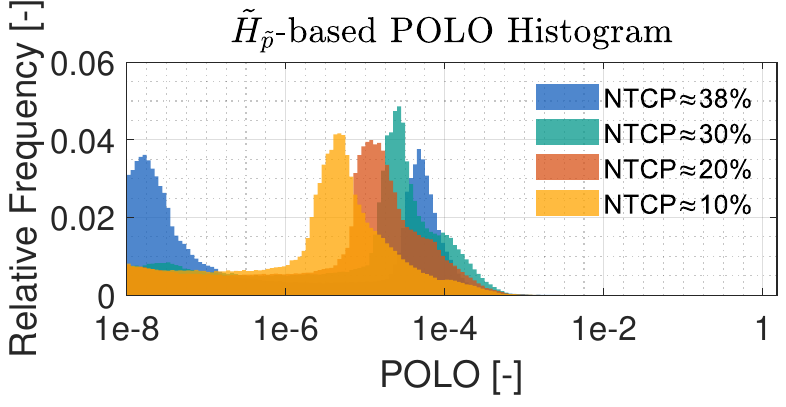} \\[-0.2em]
\end{tabular}
    \end{center}
    \vspace{-1em}
    \caption{\footnotesize\textbf{Optimal POLO histograms at different NTCP levels for all POLO model-based objectives.} We observe a consistent shift of the histogram mass to the left hand side, i.e., a higher importance of the POLO model correlates with lower optimal $p$ values.}
    \label{fig:polovh}
\end{figure}

Globally, we observe an analogous shift in the \enquote{mass} of the $p$ distribution to lower values from the histograms in \Cref{fig:polovh}. This indicates that the overarching goal of reducing $p$ during optimization can be fulfilled. 

\subsection{NTCP analysis} \label{subsec:outcomes}
\begin{figure}[htb]
    \begin{center}
        \setlength\tabcolsep{2pt} 
\centering
\begin{tabular}{@{} r M{0.5\linewidth} M{0.5\linewidth} @{}}
    & \includegraphics[width=1\linewidth, trim=7.7cm 7cm 8.4cm 7.25cm, clip=True]{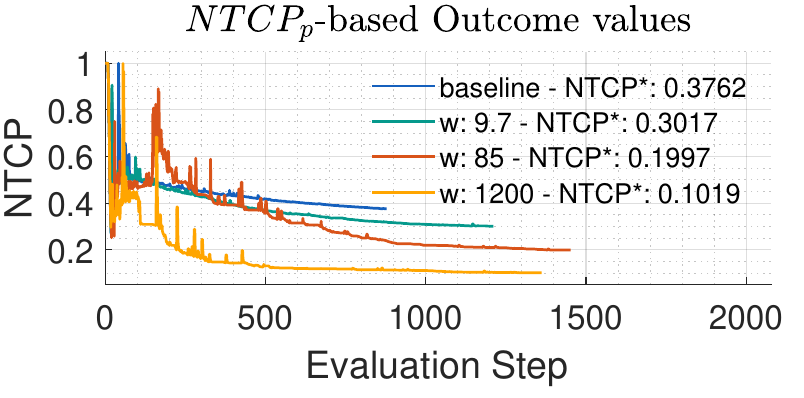}
    & \includegraphics[width=1\linewidth, trim=7.7cm 7cm 8.4cm 7.25cm, clip=True]{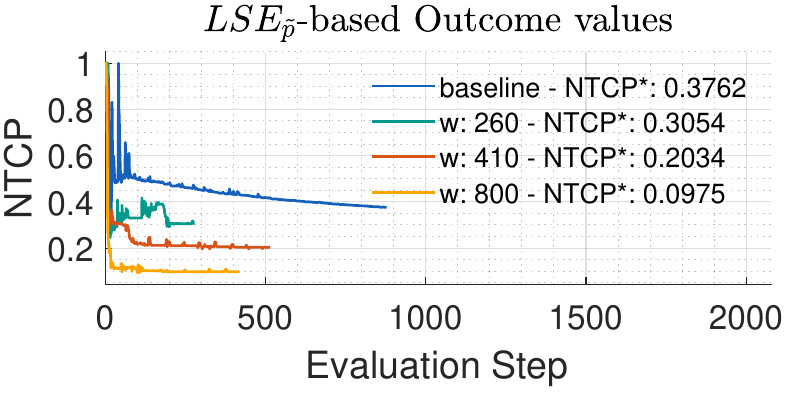} \\[-0.2em]
    & \includegraphics[width=1\linewidth, trim=7.7cm 7cm 8.4cm 7.25cm, clip=True]{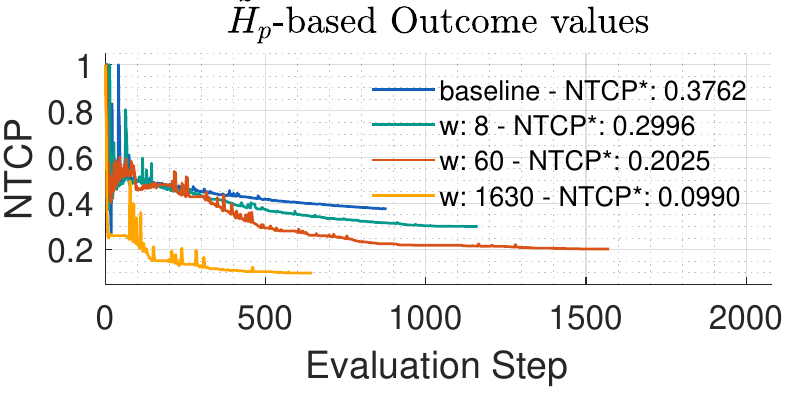}
    & \includegraphics[width=1\linewidth, trim=7.7cm 7cm 8.4cm 7.25cm, clip=True]{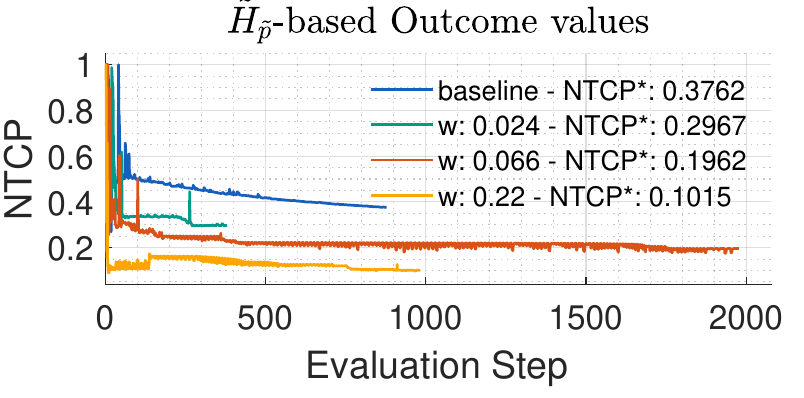} \\[-0.2em]
\end{tabular}
    \end{center}
    \vspace{-1em}
    \caption{\footnotesize\textbf{NTCP curves for the baseline and optimal plans at different NTCP levels $w$ of the POLO model-based objectives.} All objectives achieve the desired NTCP levels, but differ in terms of weight parameters, stability and convergence. Broadly speaking, the weights depend on the properties of the respective function, and those based on the linear reformulation from \Cref{eq:polo-surr} tend to converge faster.}
    \label{fig:ntcp-graphs}
\end{figure}

We optimized the treatment plan from \cref{subsec:slices} to NTCP levels of $29.7-30.5$ \si{\percent}, $19.6-20.3$ \si{\percent} and $9.8-10.2$ \si{\percent}, and while the optimal slice images and histograms show the effect of the POLO model, they are not indicative of the stability and convergence behavior of the outcome predictions derived from each optimization function (by evaluating the iterative POLO map with $NTCP_{p}$). \Cref{fig:ntcp-graphs} therefore plots the NTCP estimates over the evaluation steps of the optimization scenarios.

First, we note that the scaling of the weight parameter strongly depends on the optimization function, and is not proportional to the NTCP. It seems that $LSE_{\tilde{p}}$ responds slowly until the NTCP reduction, but requires a smaller weight multiplier to reach lower NTCP levels than the other functions. $NTCP_{p}$ and $\tilde{H}_{p}$ score the weights in the same order of magnitude, but need to be more and more \enquote{pushed} to achieve lower NTCP values. $\tilde{H}_{\tilde{p}}$ has smaller weights due to the larger function values, but for this, lower NTCP levels are obtained with a rather constant weight multiplier.

Second, we see differences in how smoothly and efficiently the optimization functions approach the respective NTCP level. Initially, $NTCP_{p}$ and $\tilde{H}_{p}$ fluctuate more widely, but then level off after around a third of the iterations. Analyzing the graphs of the optimization functions based on the linear reformulation ($LSE_{\tilde{p}}$ and $\tilde{H}_{\tilde{p}}$), we find a tendency towards faster convergence rates, but also stagnations on plateaus with the need for local line searches, which could be explained by gradient overshooting.


\section{Discussion} \label{sec:discussion}
This work introduces a technical framework for direct optimization of the probability of lesion origin in intensity-modulated proton treatment planning for LGG patients. We implemented the forward and backward calculation steps according to \Cref{fig:calc-tree} and proposed four different optimization functions ($NTCP_{p}$, $LSE_{\tilde{p}}$, $\tilde{H}_{p}$ and $\tilde{H}_{\tilde{p}}$), which either scalarize the mapping from an extended version of the original POLO model $p$ or from the linear reformulation $\tilde{p}$.

By mimicking the proton treatment plan of a LGG sample patient in matRad, we were able to show the feasibility and overall efficacy of POLO model-based optimization over a number of optimization scenarios with prescribed NTCP levels. 

Finally, we observed expectable shifts in dose, dose-averaged LET and POLO distributions across the optimization functions and found that an NTCP reduction by more than \SI{27}{\pp} is achievable for the sample patient without compromising target coverage. Nonetheless, there are several aspects to be discussed which we will address in the following.

\subsection{Dose and LET calculation}
For the purpose of this proof-of-concept, we calculate the dose-influence matrix $\mathcal{D}$ and the LET\textsubscript{d}-influence matrix $\mathcal{L}$ with pencil-beam kernel algorithms. These are known to be very fast, but rather inaccurate compared to, e.g., Markov chain Monte Carlo (MCMC) simulations. We accepted this state of affairs in view of the fact that this paper only represents a feasibility study on POLO model-based treatment plan optimization using a gradient-based framework, and in view of the time saved when running through the optimization scenarios. However, we are planning to fall back on more accurate calculation algorithms in multi-patient follow-up studies to avoid artifacts in the optimal dose and LET\textsubscript{d} distributions.

\subsection{Linear reformulation of the POLO model}
In \cref{subsec:functions}, we emphasized the non-ideal optimization properties of the original POLO model, and therefore we introduced its linear reformulation, not as an approximation of the original model, but rather to provide a means for convex optimization and apply stronger penalties on large POLO values. While the reformulation is advantageous in theory, in practice there are potential scaling issues to be considered due to the extended value range of the decision function in \Cref{eq:polo-surr}. This leads to variants of the optimization problem from \Cref{eq:problem} with different optimal points, and we observe e.g. from \Cref{fig:optimal-dose} that target coverage may not be clinically acceptable when using the linearly reformulated POLO model. It should be noted that the upper limit of infinity for the decision function is never reached in real-world planning scenarios, since $d_{RBE,fx}$ and $l_{d}$ assume finite values, but it is likely that lower weights are necessary to balance the partial functions in the weighted sum term, as well as to avoid gradient overshooting for better convergence. 

\subsection{POLO model-based optimization functions}
The largest degree of freedom in our technical framework is the rightmost element of the forward and backward calculation tree, the POLO model-based optimization function $f_{p/\tilde{p}}$. Naturally, there is an infinite number of functional designs, some of which may target the hot spots, others the centroid, and still others a non-standard, higher-order feature of the $p$ distribution. We proposed only a small selection of technically feasible, interpretable optimization functions with variable points of attack to reduce $p$ in this paper, but even for these \enquote{simple} functions we found evidence of behavioral differences in terms of stability and convergence in the optimization, along with non-uniform weighting schemes and non-proportional relationships of the weights to the final NTCP estimates. 

Therefore, while our framework facilitates direct control of treatment planning based on the POLO model, there are important design decisions left to the users that should be customized depending on the LGG patient case.

\subsection{Mathematical and clinical perspective}
The optimization results should be evaluated both mathematically and clinically. First, we were able to demonstrate the technical feasibility of POLO model-based treatment plan optimization for a selected LGG patient. 

Second, a good agreement of the optimization results, i.e., spatial distributions of $d_{RBE,fx}$ (\Cref{fig:optimal-dose} and \ref{fig:dvh}), $l_{d}$ (\Cref{fig:optimal-let} and \ref{fig:letvh}) and $p$ (\Cref{fig:optimal-polo} and \ref{fig:polovh}), with the implications of the POLO model from \cref{subsec:polo} can be found. This manifests for example in the shifting of $d_{RBE,fx}$ and $l_{d}$ such that lower values are found proximal to the VS or in the elementary product term in \Cref{eq:orig-polo}, leading to attenuated $p$ values. But, as we stated in \Cref{subsec:functions}, the POLO model-based objective functions touch on different aspects of $p$ (hence also $d_{RBE,fx}$ and $l_{d}$), and due to the degeneracy of the solution space for specific NTCP values obtained from the $NTCP_{p}$ model in \Cref{eq:ntcp}, it occurs that differences in the concrete results can be observed, for instance, when trading off dose and dose-averaged LET.

Third, the outcome predictions (\Cref{fig:ntcp-graphs}) seem to converge across all scenarios, which indicates well-posed optimization problems. However, the choice of weights associated with the objective functions is crucial for optimization success, and we found that this is not always intuitive and might follow a trial-and-error approach. Considering the varying scales of dose-volume and POLO model-based objective functions, more sophisticated weighting schemes could come into play. Beyond that, although technically feasible, we intentionally omitted further extensions of the optimization method, e.g. robust optimization, as none of these were applied in the clinical reference plan for the LGG patient, and their inclusion would limit comparability of our results with those of the ongoing phase II trial and the POLO model integration strategy developed by Sallem et al.\cite{Sallem2024}. To demonstrate feasibility, however, we added results from a composite worst-case robust optimization in the \hyperref[subsec:robust-opt]{\textbf{Appendix}}.

With regard to clinical application, the integration framework offers the user some flexibility in terms of personalized treatment planning by allowing patient-specific adjustments regarding individual anatomical variations, tumor topography relative to the ventricles, and desired dose levels. At the same time, the impact of the POLO model can be adjusted via the weight of the respective objective function, enabling the clinician to somewhat control the trade-off between dose, dose-averaged LET and NTCP reduction. This approach is akin to the work of Sallem et al.\cite{Sallem2024}, who showed that clinical objectives can be achieved while reducing NTCP levels for a number of LGG patients. Also, the INDIGO (INDIvidualized, model-Guided Optimization of proton beam treatment plans for LGG treatment) protocol presented there promotes active CTV dose de-escalation, thus we did not strictly enforce target coverage in our approach to observe the resulting trade-offs. For method selection, the capability to preserve target coverage may, however, serve as one of the decision criteria.

Nonetheless, we analyzed our integration framework by picking a single LGG patient, which is sufficient to show the feasibility and potential implications of the approach, but limits the ability to draw conclusions about the generalizability of the framework. In this regard, we are not expecting technical issues when transferring the direct optimization approach to other patients. However, uncertainties about the clinical viability would be best addressed by a full treatment plan comparison study on a larger dataset, preferably on the patient cohort currently being recruited for the ongoing phase II trial. A future study such as this could focus on the more clinical aspects of POLO model-based treatment planning and on the reproducibility of clinical decision making in the ongoing trial.

Adding more complex tumor geometries and more intricately located OARs could render the integration framework even more comprehensive. Beyond this, a comparative analysis with the methodology of Sallem et al. \cite{Sallem2024} could provide crucial insights into the clinical feasibility and potential advantages of our approach in real-world settings, which is essential for evaluating its clinical impact and practical implementation. Specifically, comparing the dose distributions between our study and that of Sallem et al. \cite{Sallem2024} at equivalent NTCP levels would yield valuable information on target coverage and the sparing of OARs, with a particular focus on the cerebral ventricles.

\subsection{Translation to other models}
The methodology described in \cref{sec:methods} can be transferred to other voxel-based outcome models if (i) the model assumes a dose-response relationship, i.e., takes into account dosimetric variables, (ii) the model provides a gradient with respect to the dosimetric variables, and (iii) the voxel-wise model output can be scalarized with an optimization function $f$.

If these criteria are met, implementing the outcome model for treatment plan optimization is straightforward and only requires to set up the optimization function (e.g. to transform $p(\eta, k)$ into $f_{p}$), as well as to define the projection between the spatial dose and outcome distributions (e.g. to calculate $p(\eta, k)$ from $\phi$ forwards and derive $\nabla_{\phi}{f_{p}}$ backwards). This holds true for a wide range of statistical and machine learning models, including linear models and (deep) neural networks.


\section{Conclusion} \label{sec:conclusion}
We presented a technical framework to directly integrate the POLO model from Bahn et al. \cite{Bahn2020} into proton treatment planning for LGG patients by defining the forward calculation and backward differentiation steps required to perform model-based dose optimization. 

Implementing feasible optimization functions and running a number of optimization scenarios for fixed NTCP levels, we were able to generate results on the spatial $d_{RBE,fx}$, $l_{d}$ and $p$ distributions. From a mathematical standpoint, the results are consistent regarding changes expected from model inspection and changes observed between plans with different weights of the POLO model-based optimization function, and we also observe a consistent reduction in NTCP across all scenarios without affecting target coverage for a sample patient.

As already mentioned, the clinical feasibility of the integration framework, particularly its capacity to balance effective tumor control with the minimization of side effects, still requires comprehensive evaluation to ensure that it is both practical and beneficial in a real-world clinical setting. Conclusively, the integration framework may well contribute to automated POLO model-based risk minimization in proton treatment planning for LGG patients, and should further be evaluated in multi-patient follow-up studies.


\section*{Acknowledgements}
The present contribution is supported by the Helmholtz Association under the joint research school HIDSS4Health — Helmholtz Information and Data Science School for Health.
HS and JB acknowledge financial support from of the \enquote{Alois-Hirdt-Erben und Wieland} foundation Heidelberg (Germany). NW acknowledges funding by the Deutsche Forschungsgemeinschaft
(DFG, German Research Foundation) – Project No.~443188743.


\section*{Conflict of Interest Statement}
The authors have no relevant conflicts of interest to disclose.


\section*{ORCID iDs}
Tim Ortkamp \orcidlink{0009-0005-6603-3088}, Habiba Sallem \orcidlink{0009-0004-7811-7839}, Semi Harrabi \orcidlink{0000-0003-1195-7150}, 
Oliver Jäkel \orcidlink{0000-0002-6056-9747}, Julia Bauer \orcidlink{0000-0002-7944-3757}, Niklas Wahl \orcidlink{0000-0002-1451-223X}


\printbibliography


\appendix
\label{appendix}
\section*{Appendix} \label{sec:matrad-functions}
Let $\tilde{d}=\mathcal{D}_{r}\phi\in{\mathbb{R}_{+}^{n_{r}}}$ be the vectorized dose distribution in the volume of interest $r$, using the dose-influence matrix $\mathcal{D}\in{\mathbb{R}_{+}^{n\times{m}}}$ and the fluence vector $\phi\in{\mathbb{R}_{+}^{m}}$. Then, the dose optimization functions stated in \cref{subsec:design} are:

\subsection*{Dose Uniformity}
Dose uniformity measures the standard deviation of the dose distribution, i.e.,
\begin{align*}
    f_{\text{Dose Uniformity}}(\tilde{d}) = \sqrt{\mathds{V}ar(\tilde{d})}\,.
\end{align*}

\subsection*{Squared Deviation}
Squared deviation includes a reference dose parameter $d^{ref}$, and quantifies the normalized sum of squared differences between the dose values and the reference value, i.e.,
\begin{align*}
    f_{\text{Squared Deviation}}(\tilde{d}) = \frac{1}{n_{r}}(\tilde{d}-d^{ref})^{T}(\tilde{d}-d^{ref})\,.
\end{align*}

\subsection*{Squared Underdosing}
Squared underdosing is structurally similar to squared deviation, except that it uses a minimum dose parameter $d^{min}$ and only penalizes lower values, yielding the function
\begin{align*}
    f_{\text{Squared Underdosing}}(\tilde{d}) = \frac{1}{n_{r}}\varepsilon^{T}\varepsilon\,,\quad\varepsilon_{i}=\begin{cases}
        d^{min}-\tilde{d}_{i}\,,\quad &\tilde{d}_{i}<{d^{min}} \\
        0\,,\quad &\text{else}
    \end{cases}\,,
\end{align*}
where $\varepsilon$ \enquote{clamps} the dose difference values to only consider underdosed voxels.

\subsection*{Maximum DVH}
Maximum DVH takes a reference dose $d^{ref}$ and a maximum volume fraction $v^{max}$ as input parameters, calculates the dose quantile $d^{v}$ at $v=v^{max}$ and then determines
\begin{align*}
    f_{\text{Maximum DVH}}(\tilde{d}) = \frac{1}{n_{r}}\varepsilon^{T}\varepsilon\,,\quad\varepsilon_{i}=\begin{cases}
        \tilde{d}_{i}-d^{ref}\,,\quad &\tilde{d}_{i}\in(d^{ref}, d^{v}] \text{ and } d^{ref}<d^{v} \\
        0\,,\quad &\text{else}
    \end{cases}\,,
\end{align*}
where $\varepsilon$ is only nonzero if the respective dose value contributes to moving the DVH quantile beyond the reference value.

\subsection*{Minimum DVH}
Minimum DVH, analogous to maximum DVH, utilizes the value function
\begin{align*}
    f_{\text{Minimum DVH}}(\tilde{d}) = \frac{1}{n_{r}}\varepsilon^{T}\varepsilon\,,\quad\varepsilon_{i}=\begin{cases}
        \tilde{d}_{i}-d^{ref}\,,\quad &\tilde{d}_{i}\in(d^{v}, d^{ref}] \text{ and } d^{v}<d^{ref} \\
        0\,,\quad &\text{else}
    \end{cases}\,,
\end{align*}
but defines a minimum volume fraction $v^{min}$ instead of $v^{max}$.

\subsection*{Minimum/Maximum Dose}
Minimum/Maximum dose implements a box constraint on the dose, i.e., the dose values are lower- and upper-bounded by $d^{min}$ and $d^{max}$. The constraint function is given by the identity and therefore the constraint reads
\begin{align*}
    d^{min}\leq{\tilde{d}}\leq{d^{max}}\,.
\end{align*}
Internally, matRad translates such a constraint by applying the log-sum-exp approximation.

\subsection*{Computational costs}
\Cref{tab:runtimes} lists the number of IPOPT evaluations, CPU and wall clock runtimes for optimizing the baseline and POLO model-based treatment plans from \Cref{subsec:design}. 
\begin{table}[htb]
    \centering \renewcommand{\arraystretch}{1}
\setlength\tabcolsep{3pt}
{\fontsize{9}{9} \selectfont
\begin{tabular}{ccccc}
        \textbf{Optimization case} & \textbf{\# IPOPT evaluations} & \textbf{CPU time [\si{\second}]} & \textbf{Wall clock time [\si{\second}]} \\ \toprule
        Baseline ($NTCP=37.6\%$) & 878 & 82 & 128 \\[1em]

        $+NTCP_{p}$ (30.2\%) & 1211 & 719 & 771 \\
        $+NTCP_{p}$ (20.0\%) & 1452 & 665 & 711 \\
        $+NTCP_{p}$ (10.2\%) & 1362 & 584 & 627 \\[1em]

        $+LSE_{\tilde{p}}$ (30.5\%) & 279 & 87 & 120 \\
        $+LSE_{\tilde{p}}$ (20.3\%) & 515 & 141 & 175 \\
        $+LSE_{\tilde{p}}$ (9.8\%)\phantom{0} & 420 & 120 & 152 \\[1em]

        $+\tilde{H}_{p}$ (30.0\%) & 1162 & 675 & 726 \\
        $+\tilde{H}_{p}$ (20.3\%) & 1572 & 662 & 706 \\
        $+\tilde{H}_{p}$ (9.9\%)\phantom{0} & 646 & 289 & 326 \\[1em]

        $+\tilde{H}_{\tilde{p}}$ (29.7\%) & 381 & 114 & 147 \\
        $+\tilde{H}_{\tilde{p}}$ (19.6\%) & 1977 & 532 & 569 \\
        $+\tilde{H}_{\tilde{p}}$ (10.2\%) & 983 & 264 & 298 \\\bottomrule

    \end{tabular}}
    \caption{\footnotesize\textbf{Optimization runtimes in matRad for baseline and POLO model-based treatment plans.} The baseline plan requires \SI{0.14}{\second} (CPU) or \SI{0.35}{\second} (wall clock) per iteration, excluding LET and POLO calculation. By adding the POLO model-based optimization components as described in \Cref{sec:methods}, time per evaluation increases by a factor of 2.9 -- 6.4 (CPU) or 2.0 -- 4.4 (wall clock).}
    \label{tab:runtimes}
\end{table}

\subsection*{Robust optimization} \label{subsec:robust-opt}
\begin{figure}[htb]
    \begin{center}
        \setlength\tabcolsep{0pt} 
\begin{tabular}{@{} r M{0.333\linewidth} M{0.333\linewidth} M{0.333\linewidth} @{}}
  & $d_{RBE,fx}$\phantom{ABCD} & $l_{d}$\phantom{ABCD} & $p$\phantom{ABCD} \\
  \rotatebox[origin=c]{90}{\hspace{5pt}Baseline}
  & \includegraphics[width=1\linewidth, trim=4.7cm 10.2cm 2.6cm 9.9cm, clip=True]{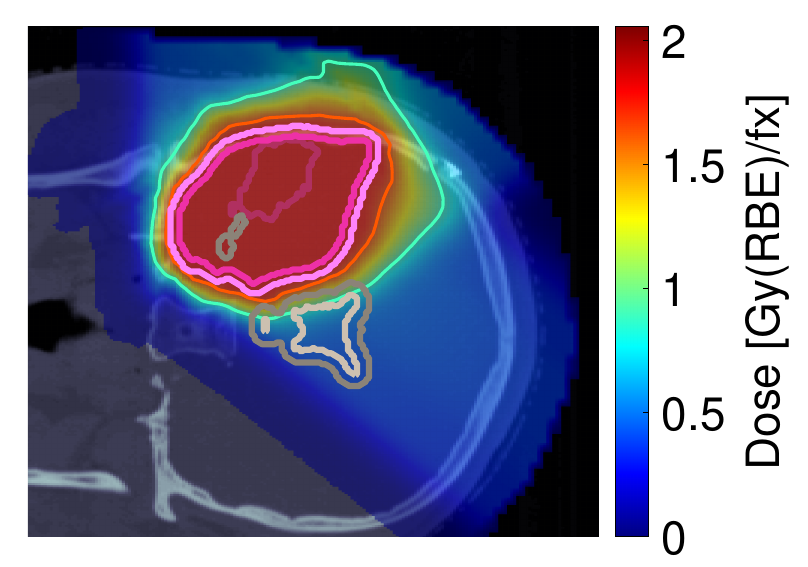} 
  & \includegraphics[width=1\linewidth, trim=4.7cm 10.2cm 2.6cm 9.9cm, clip=True]{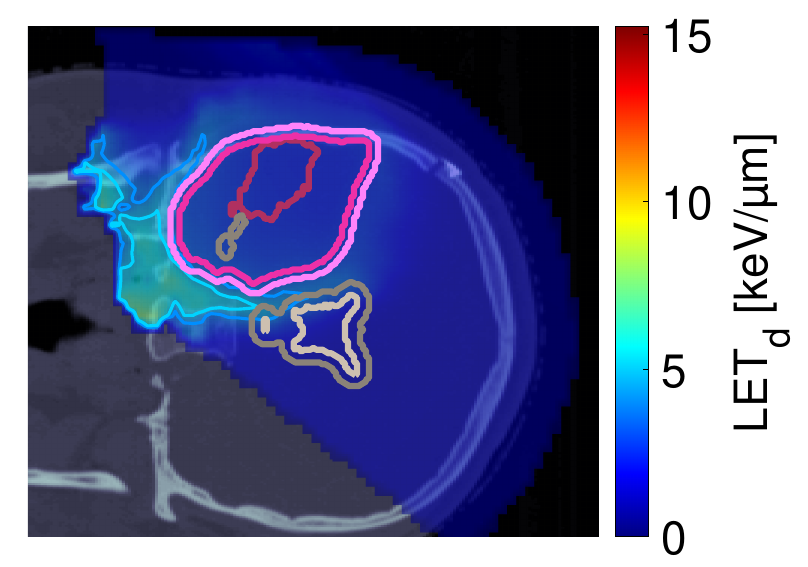} 
  & \includegraphics[width=1\linewidth, trim=4.7cm 10.2cm 2.6cm 9.9cm, clip=True]{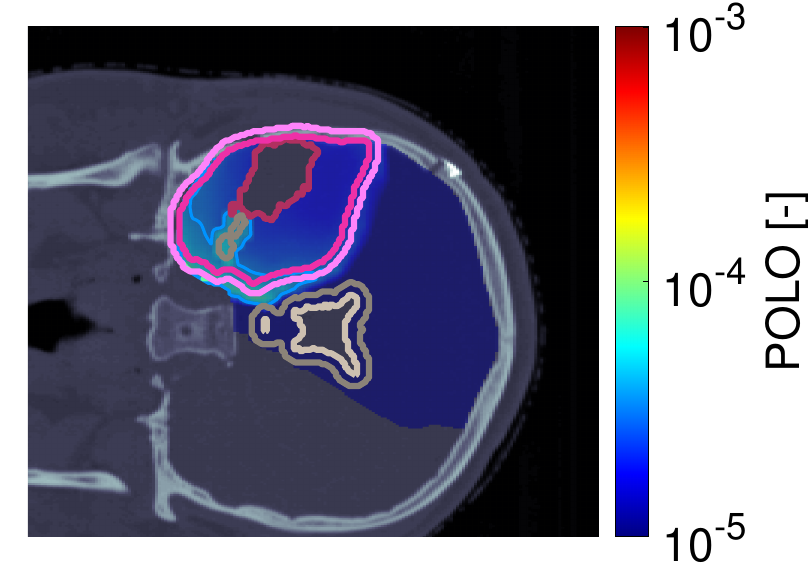} \\
  \rotatebox[origin=c]{90}{\hspace{5pt}POLO}
  & \includegraphics[width=1\linewidth, trim=4.7cm 10.2cm 2.6cm 9.9cm, clip=True]{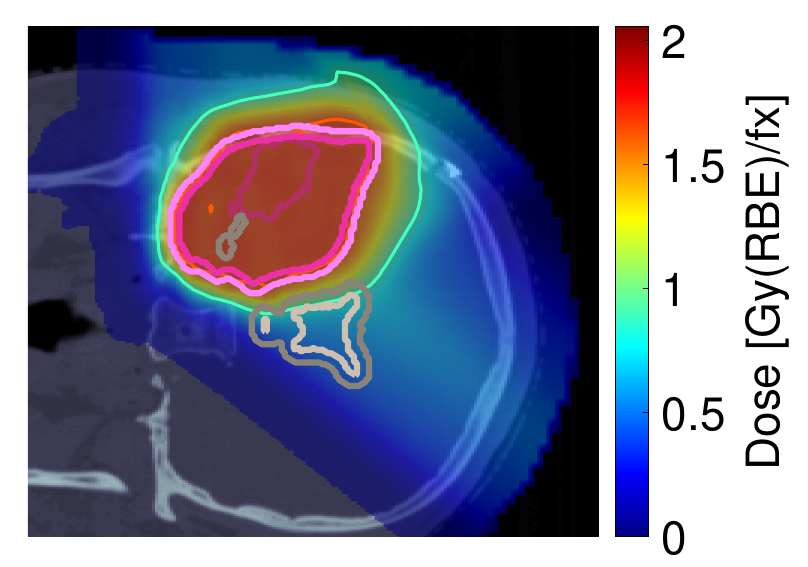} 
  & \includegraphics[width=1\linewidth, trim=4.7cm 10.2cm 2.6cm 9.9cm, clip=True]{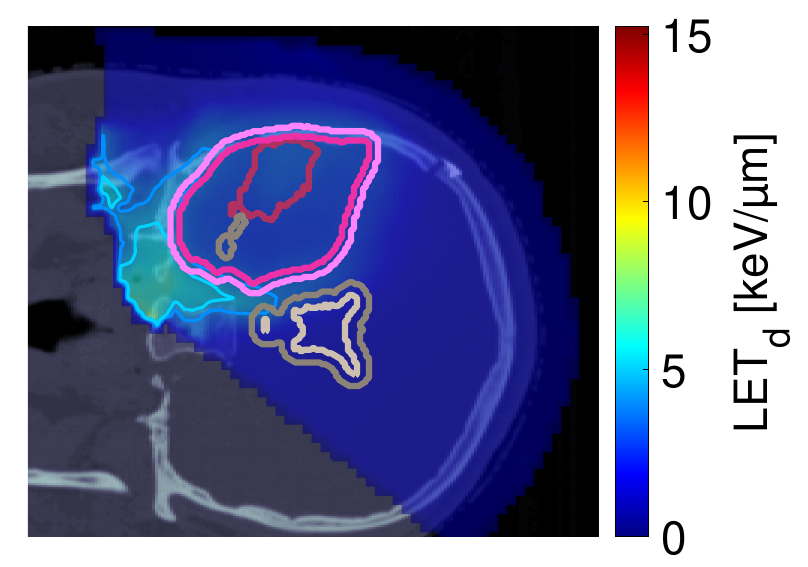} 
  & \includegraphics[width=1\linewidth, trim=4.7cm 10.2cm 2.6cm 9.9cm, clip=True]{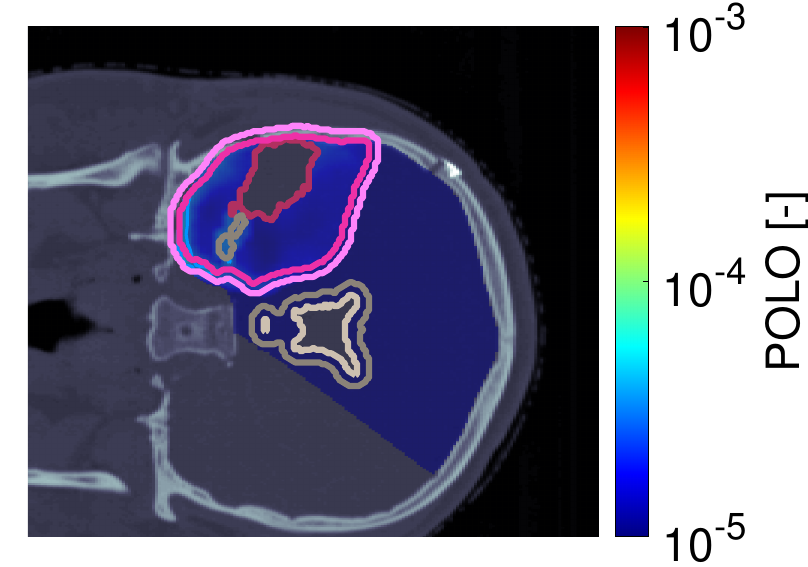} \\
  \vspace{-20pt}
\end{tabular}
\par\vspace{1em}
\begin{tabular}{@{} r M{1\linewidth} M{1\linewidth} @{}}
    & \includegraphics[width=1\linewidth, trim=2.2cm 7cm 2.9cm 7.1cm, clip=True]{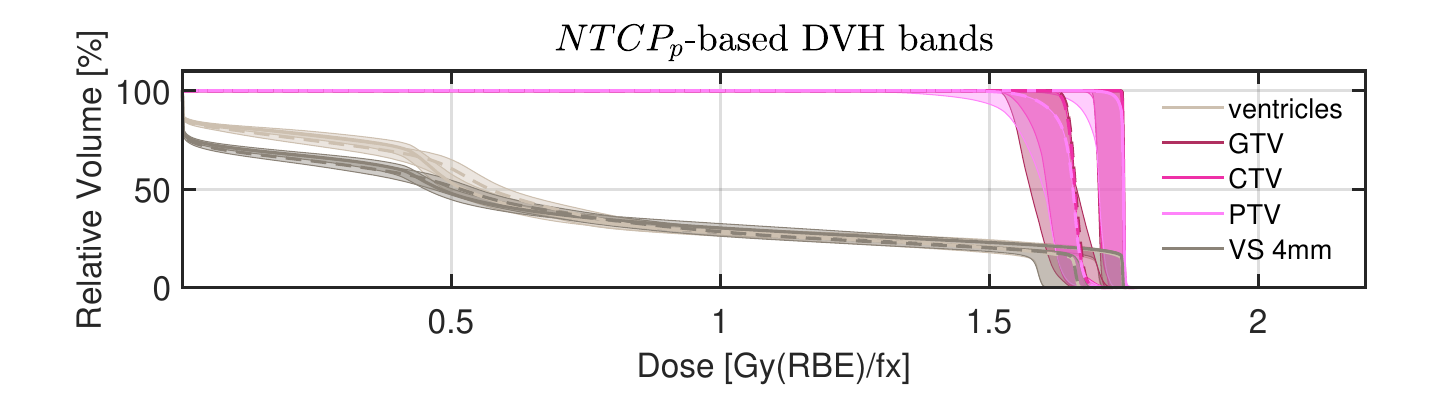} \\[-0.2em]
    & \includegraphics[width=1\linewidth, trim=2.2cm 7cm 2.63cm 7.1cm, clip=True]{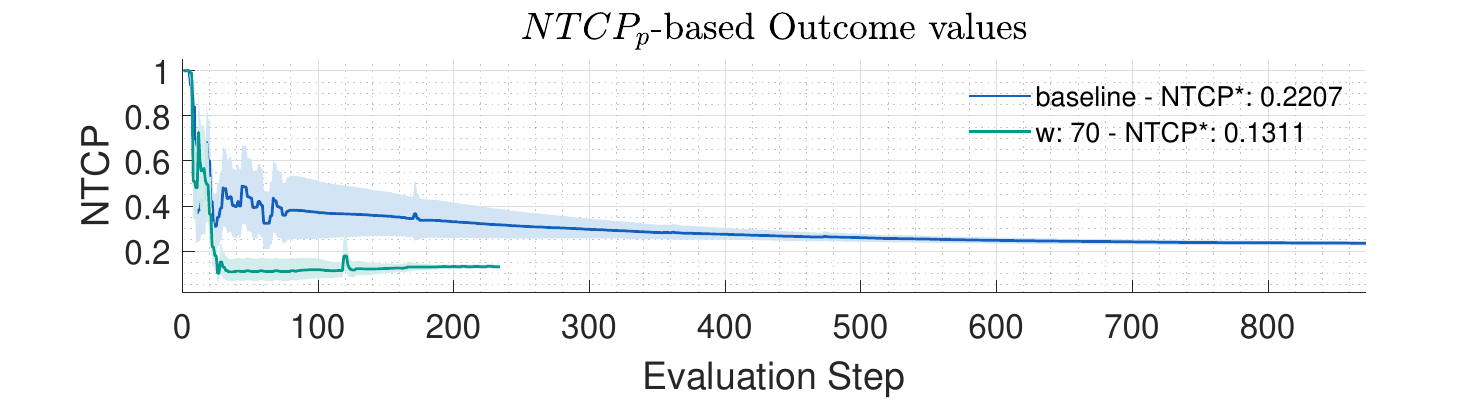} \\[-2em]
\end{tabular}
    \end{center}
    \caption{\footnotesize\textbf{Composite worst-case robust optimization results.} Top panels show slice images for dose, LET\textsubscript{d} and POLO. Bottom panels display the respective DVHs (baseline: \protect\solid, POLO: \protect\dashed) and NTCP curves (including scenario bands). We observe a dose reduction and LET redistribution, leading to an overall reduction in the nominal NTCP from \SI{22}{\percent} (baseline) to \SI{13}{\percent} (POLO model-based).}
    \label{fig:robust-opt-results}
\end{figure}

To address clinical requirements for plan stability, we also performed robust POLO model-based treatment plan optimization. The uncertainty set comprises nine discrete scenarios (eight perturbed and one nominal), incorporating $\pm{2.5} \si{\percent}$ range and 2 \si{\milli\meter} setup uncertainty. We employed a composite worst-case approach; however, dose uniformity objectives were kept nominal to prevent excessive \enquote{dose smearing}. As proof-of-concept, we integrated the POLO model via $NTCP_{p}$. \Cref{fig:robust-opt-results} illustrates the optimization results for the robust baseline plan (nominal NTCP: \SI{22}{\percent}) and the robust POLO model-based plan (nominal NTCP: 13 \si{\percent}), including slice images and DVH/NTCP uncertainty bands. Our results indicate reduced maximum doses to the tumor volumes and the VS, alongside an LET shift away from the latter. Although NTCP decreased, earlier optimizer termination suggests objective competition and the attainment of a Pareto-optimal trade-off between tumor coverage and OAR sparing. More detailed investigation, however, goes beyond the scope of this paper, especially since the clinical reference protocol does not include robustness. 

\end{document}